%% file: main.tex
\documentclass[10pt,journal,compsoc]{IEEEtran}

\usepackage{paralist}

\UseRawInputEncoding

\ifCLASSOPTIONcompsoc
  \usepackage[nocompress]{cite}
\else
  \usepackage{cite}
\fi

\ifCLASSINFOpdf
   \usepackage[pdftex]{graphicx}
\else
   \usepackage[dvips]{graphicx}
\fi

\usepackage{amsmath}
\usepackage{amssymb}

\usepackage{array}

\usepackage{xcolor}  
\newcommand{\result}[1]{
\begin{center}
    \fcolorbox{black}{gray!10}{\parbox{.96\columnwidth}{#1}}
\end{center}
}

\usepackage{caption}
\usepackage{subcaption}

\usepackage{url}

\usepackage{xspace}
\usepackage{setspace}
\newcommand{\uncertum}{{\textit{UncerTum}}\xspace}
\newcommand{\uncertest}{{\textit{UncerTest}}\xspace}
\newcommand{\probFirst}{{\textit{Prob.1 f(PET,PTR,AUM)}}\xspace}
\newcommand{\probSecond}{{\textit{Prob.2 f(PET,PTR,PUS)}}\xspace}
\newcommand{\probThird}{{\textit{Prob.3 f(PET,PTR,ANU)}}\xspace}
\newcommand{\probFourth}{{\textit{Prob.4 f(PET,PTR,PUU)}}\xspace}
\newcommand{\probFifth}{{\textit{Prob.5 f(PET,PTR,AUM,PUS)}}\xspace}
\newcommand{\probSixth}{{\textit{Prob.6 f(PET,PTR,AUM,ANU)}}\xspace}
\newcommand{\probSeventh}{{\textit{Prob.7 f(PET,PTR,AUM,PUU)}}\xspace}
\newcommand{\probEighth}{{\textit{Prob.8 f(PET,PTR,PUS,ANU)}}\xspace}
\newcommand{\probNineth}{{\textit{Prob.9 f(PET,PTR,PUS,PUU)}}\xspace}
\newcommand{\probTenth}{{\textit{Prob.10 f(PET,PTR,ANU,PUU)}}\xspace}
\newcommand{\aum}{{\textit{AUM}}\xspace}
\newcommand{\anu}{{\textit{ANU}}\xspace}
\newcommand{\nsu}{{\textit{NU}}\xspace}
\newcommand{\anou}{{\textit{ANOU}}\xspace}
\newcommand{\nou}{{\textit{NOU}}\xspace}
\newcommand{\um}{{\textit{UM}}\xspace}
\newcommand{\pus}{{\textit{PUS}}\xspace}
\newcommand{\puu}{{\textit{PUU}}\xspace}
\newcommand{\usp}{{\textit{USP}}\xspace}
\newcommand{\nsuu}{{\textit{NUU}}\xspace}
\newcommand{\uncertaintyaware}{{uncertainty-aware}\xspace}
\newcommand{\MOSA}{{\revision{MuOSA}}\xspace}
\newcommand{\approach}{{\revision{UncerPrio}}\xspace}

\usepackage{threeparttable}
\usepackage{multirow}
\usepackage{color}
\usepackage{colortbl}
\usepackage{pgfplots}
\usepackage{comment}
\usepackage{algorithm}
\usepackage{algpseudocode}
\usepackage{booktabs}
\usepackage{bm}
\usepackage{comment}

\newcommand{\Atwelve}{\^{A}\textsubscript{12}\xspace}

\definecolor{greenM}{RGB}{240,255,240}
\definecolor{blueS}{RGB}{240,248,255}
\definecolor{pinkN}{RGB}{255,235,235}
\definecolor{purpleC}{RGB}{230,225,250}
\definecolor{multi}{RGB}{255,255,240}

\newcommand{\revision}[1]{\color{black}{#1}\color{black}\xspace}

\begin{document}
%
\title{Uncertainty-Aware Test Prioritization: Approaches and Empirical Evaluation}

\author{Man~Zhang,
        Jiahui~Wu,
        Shaukat~Ali
        and Tao~Yue

\IEEEcompsocitemizethanks{\IEEEcompsocthanksitem M. Zhang is with Kristiania University College, Oslo, 999026, Norway.\protect\\
E-mail: man.zhang@kristiania.no
\IEEEcompsocthanksitem J. Wu, S. Ali, and T. Yue are with Simula Research Laboratory, Oslo, 999026, Norway. 
E-mail: \{jiahui, shaukat, tao\}@simula.no}
}

%
%

\markboth{Journal of \LaTeX\ Class Files,~Vol.~14, No.~8, August~2015}%
{Shell \MakeLowercase{\textit{et al.}}: Uncertainty Wise and Time Aware Test Case Prioritization}
%



\IEEEtitleabstractindextext{%
\begin{abstract}
Complex software systems, e.g., Cyber-Physical Systems (CPSs), interact with the real world; thus, they often behave unexpectedly in uncertain environments. \revision{Testing such systems is challenging due to limited resources, time, and complex testing infrastructures setup. Furthermore, the inherent uncertainties in their operating environment complicate their testing.} 
Devising \textit{\uncertaintyaware testing} solutions supported with test optimization techniques (e.g., with search) can be considered as a mandate for tackling this challenge. This paper proposes an \textit{\uncertaintyaware} and \textit{time-aware} test case prioritization approach, named \approach, for optimizing a sequence of tests to execute with a multi-objective search. 
To guide the prioritization with \textit{uncertainty}, we identify four uncertainty measures (denoted as \aum, \pus, \anu, and \puu), which characterize uncertainty measurement, uncertainty space, the number of uncertainties, and uncertainty coverage. 
Based on these measures and their combinations, we proposed 10 \uncertaintyaware and multi-objective test case prioritization problems, and each problem was additionally defined with one cost objective to be minimized (denoted as \textit{PET} for execution cost) and one effective measure to be maximized (denoted as \textit{PTR} for model coverage). 
Moreover, considering time constraints for test executions (i.e., time-aware), we defined 10 time budgets for all the 10 problems \revision{for identifying the best strategy in solving uncertainty-aware test prioritization}. In our empirical study, we employed four well-known Multi-Objective Search Algorithms (\MOSA{s}): NSGA-II, MOCell, SPEA2, and CellDE with five use cases from two industrial CPS subject systems, and used Random Algorithm (RS) as the comparison baseline. Results show that all the \MOSA{s} significantly outperformed RS.
\revision{The strategy of \probSixth} (i.e., the problem with \revision{uncertainty measures} \aum and \anu combined) \revision{
achieved the overall best performance in observing uncertainty}
when using 100\% time budget.
\end{abstract}

}

\maketitle

\IEEEdisplaynontitleabstractindextext

%
\IEEEpeerreviewmaketitle


\IEEEraisesectionheading{\section{Introduction}\label{sec:introduction}}

\IEEEPARstart{A}{long} with the increase in the number of the deployment of large-scale and complex software systems such as Cyber-Physical Systems (CPSs), growing attention has been paid to their dependability when facing uncertainty during their operations, as such systems often interact with the real world, which itself is inherently uncertain. 
Testing has been a primary means to ensure the dependability of such systems before their deployment~\cite{patton2006software}. 

In any non-trivial context, test case prioritization is essential because test execution is both time-wise and monetary expensive, especially when testing complex software systems such as CPSs, which often requires the use of test infrastructures (e.g., simulators, emulators, and even hardware equipment) if available. Various search-based solutions have been proposed for test case prioritization in the context of CPS testing~\cite{wang2016enhancing,Resource-allocationSBST}. \revision{In addition, with the growing realization of handling inherent uncertainty in CPSs to ensure their dependability, several uncertainty-wise testing approaches have been proposed in the literature~\cite{wang2018oracles, shin2018test, menghi2019generating, shin2021uncertainty, camilli2021uncertainty}. Along these lines, in our previous work, we proposed \uncertest for uncertainty-wise test case generation and minimization for CPSs. However, we realize that such} test case generation methods (e.g., \cite{zhang2019uncertainty}) generate many test cases when considering uncertainty, and test cases must be executed multiple times \revision{since a known uncertainty might not always occur, and an unknown uncertainty might be observed.}
\revision{For instance, our previous study~\cite{zhang2019uncertainty} showed that executing more than 1000 test cases could take more than one day. Executing them multiple times due to uncertainty easily becomes infeasible in practice; therefore, in practice, test engineers need a solution to prioritize such test executions to execute \textit{effective} (e.g., uncertainty-aware in our context) tests earlier to observe \textit{more} uncertainties \textit{earlier} within a limited time budget. 
}

In this paper, we define 10 \uncertaintyaware and time-\revision{aware} test case prioritization problems by considering uncertainty, time budget, execution time, and behavior coverage, which are formulated as 10 search problems.  
Specifically, the 10 problems are defined based on the following four uncertainty, one cost, and one effectiveness measure: 
\begin{inparaenum}[(1)]
\item \textit{uncertainty measures}:
\textbf{\textit{AUM}}: the average uncertainty measure (adopted from Uncertainty Theory~\cite{liu2007uncertainty}) of the prioritized test cases (to maximize); \textbf{\textit{PUS}}: the average percentage of uncertainty space (adopted from Uncertainty Theory~\cite{liu2007uncertainty}) covered (to maximize); \textbf{\textit{ANU}}: the average number of modeled uncertainties covered (to maximize); and \textbf{\textit{PUU}}: the average number of unique uncertainties covered (to maximize);
\item \textit{cost measure}: \textbf{\textit{PET}}: the total execution time of prioritized test cases (to minimize); and
\item \textit{effectiveness measure}: \textbf{\textit{PTR}}: the (state machine) transition coverage (to maximize).
\end{inparaenum}
Each search problem comprises the cost, effectiveness, and one or two uncertainty measures, which lead to 10 fitness functions to guide search algorithms toward finding optimal solutions. In addition, for each problem, we conduct experiments with 10 levels of time budget, which are defined as 10 time constraints.

In our empirical study, we employed four well-known, multi-objective search algorithms (\MOSA{s}) (i.e., NSGA-II~\cite{deb2002fast}, MOCell~\cite{nebro2007design, nebro2009mocell}, SPEA2~\cite{zitzler2001spea2}, and CellDE~\cite{durillo2008solving}) to solve the 10 \uncertaintyaware test case prioritization problems under the 10 time constraints. We evaluated their performance with Random Search (RS), with five use cases from two industrial CPS subject systems. 
Results showed that all the selected \MOSA{s} performed significantly better than \textit{RS}. 
When comparing the \MOSA{s}, different algorithms performed best to solve different prioritization problems for different subject systems. 
Therefore, based on the results, we derived recommendations to select \MOSA{s} for solving a particular test prioritization problem for a given time budget. 
To further investigate the impacts of time budgets and compare performances achieved by various problems \revision{to identify the best strategy}, we conducted analyses on them using a metric named \anou representing efficiency in observing uncertainties with prioritized solutions.
Results of the analyses indicate that there exist significant correlations between time budget and \anou, but the direction of the correlations (i.e., negative or positive) might depend on use cases.
In this empirical study, we found that four out of the five use cases demonstrated a positive correlation; thus, 100\% time constraint (denoted as TB100) is recommended by default.
Among the 10 uncertainty-wise problems, the problems containing \aum and \anou demonstrate better performances than others. 
This hints that the two subjective measures might have a high chance of positively triggering the occurrence of uncertainties during test execution.
In the five use cases, a multi-objectives problem defined with both \aum and \anu, denoted as \probSixth, achieved the overall best performance. 

\revision{\textit{Contributions.} We: 
1) proposed an uncertainty-aware and time-aware test prioritization approach with tool support\footnote{\label{foot:link}https://github.com/man-zhang/uncertainty-prioritization} for optimizing test case execution cost-effectively for complex software systems (e.g., CPSs) which need to deal with inherent uncertainty;
2) reformulated uncertainty-aware and time-aware test prioritization as 10 multi-objective search problems (MuOSP) under 10 time constraints by considering four uncertainty-aware metrics, one cost measure, and one effectiveness measure; 
3) conducted an empirical study for 10 uncertainty-aware MuOSPs under 10 time constraints using four \MOSA{s} and RS on five use cases from two industrial systems; and
4) provided a step-wise guide to apply \approach.
}

\textit{Paper structure}: Section~\ref{sec:background} presents the overall context and background. 
Section~\ref{sec:problems} formulates the prioritization problems. Section~\ref{sec:industrialSystem} describes the industrial subject systems. Section~\ref{sec:evaluation} presents the experiments and results. 
The related work is given in Section~\ref{sec:relatedWork} and Section~\ref{sec:conclusion} concludes the paper.


\section{Background}\label{sec:background}
This section provides the necessary background information on the source and characterization of test cases to be prioritized. Section~\ref{sec:beliefModel} introduces how to construct test ready models\revision{\footnote{\revision{Test ready models capture \textit{known} behaviors and structure of the system under test (SUT) and are \textit{ready} for enabling model-based testing (MBT), e.g., for automated test case generation.}}} with uncertainty, which are used to automatically generate test cases with uncertainty with an uncertainty-aware test case generation method (Section~\ref{sec:testCaseUncerTest}). Generated tests are then used as input by our \uncertaintyaware prioritization. 
Section~\ref{sec:testResults} introduces the uncertainty-wise test verdict for asserting uncertainty occurrences.  

\subsection{Belief test ready models developed with UncerTum}\label{sec:beliefModel}
In our previous work, we have developed an integrated, UML-based modeling framework (named as \textit{UncerTum})~\cite{zhang2019uncertainty} for constructing test ready models with \revision{\textit{subjective} uncertainty}\footnote{\revision{From a subjective perspective, Zhang et al.~\cite{zhang2016understanding} defined uncertainty as "a state of affairs whereby a \textit{BeliefAgent} does not have full confidence in a \textit{Belief} that it holds".}} information explicitly specified, called \textit{belief test ready models} (BMs). \textit{UncerTum} is equipped with specialized modeling notations, named the UML Uncertainty Profile (\textit{UUP}), for specifying uncertainties. As the core of \textit{UncerTum}, \textit{UUP} characterizes an uncertainty with information such as \textit{IndeterminacySource} (for capturing a source that may lead to the occurrence of the uncertainty) and quantifies it via \textit{Measurement}. \textit{UncerTum} also defines four sets of UML model libraries: \textit{Pattern}, \textit{Time}, \textit{Measure}, and \textit{Risk} libraries. 

\begin{figure}
\centering
\includegraphics[width=0.99\linewidth]{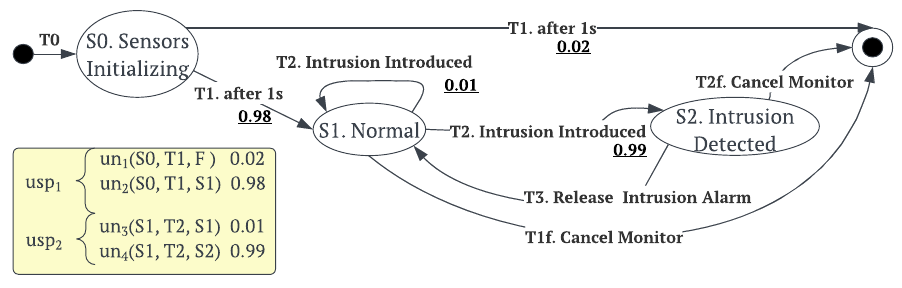}
\raggedright \footnotesize \revision{* An example of test $t_1$: \textit{I}$\rightarrow$\textit{T0}$\rightarrow$\textit{S0}$\rightarrow$\textit{T1}$\rightarrow$\textit{S1}$\rightarrow$\textit{T2}$\rightarrow$\textit{S2}$\rightarrow$\textit{T3}$\rightarrow$\textit{S1}$\rightarrow$\textit{T2} $\rightarrow$\textit{S2}$\rightarrow$\textit{T2f}$\rightarrow$\textit{F} derived with the BSM; $un_3$ links to a known indeterminacy source \textit{IndS}$_1$ to enable a motion sensor jammer.}
\vspace{-.5\baselineskip}
\caption{\revision{A simplified \textit{SafeHome} example of the belief test ready model with explicitly captured uncertainty}}
\label{fig:example}
\vspace{-1.0\baselineskip}
\end{figure}

When modeling with \textit{UncerTum}, i.e., specifying uncertainty as part of test ready models (e.g., UML state machines specifying expected system behaviours) and then forming belief test ready models, \textit{uncertainty} is a situation whereby engineers (\textit{belief agent}) lack of confidence (hence \textit{subjective}) about a system behavior of the belief agent's concern, e.g., whether a specified state occurs after invoking a specified trigger action. 
\revision{An example of a belief state machine (BSM) specified with UML state machine is shown in Figure~\ref{fig:example}, which models the \textit{Monitoring} behavior of \textit{SafeHome}~\cite{zhang2019uncertaintywise,zhang2022test} with uncertainty.
With \textit{UncerTum}, we specify one uncertainty in the BSM based on the source state, transition event, and target state. For instance, when the \textit{SafeHome} system is in state \textit{S0.Sensors Initializing} (the source state), \textit{T1.after 1s} (a transition with time event is triggered), the system might transit to any of the two potential target states (i.e., \textit{F.Final State} and \textit{S1.Normal}) non-deterministically. To capture such uncertain behaviors, we identify two uncertainties as $un_1 (S0, T1, F)$ and $un_2 (S0, T1, S1)$. 
In a BSM, we define \textit{uncertainty space}, which is the universal set of uncertainties originating from the same source target triggered by the same event, such as \textit{usp}$_1$=$\{un_1, un_2\}$. Covering an uncertainty space indicates covering a kind of uncertainty, and covering any uncertainty in an uncertainty space means that this uncertainty space is covered, e.g., \textit{usp}$_1$ is covered if any $un_1$ or $un_2$ is covered. 
In the BSM shown in Figure~\ref{fig:example}, there exist eight unique transitions\footnote{\revision{Note that a unique transition is identified by considering its source and target states, e.g., \textit{T1}$^{S0,S1}$ and \textit{T1}$^{S0,F}$ are considered as two unique transitions as their target states are different.}} (i.e., \textit{TR(BM)}=$\{$\textit{T0}$^{I,S0},$\textit{T1}$^{S0,F},$\textit{T1}$^{S0,S1},$\textit{T2}$^{S1,S1},$\textit{T2}$^{S1,S2},$\textit{T3}$^{S2,S1}$, \textit{T2f}$^{S2,F},$\textit{T1f}$^{S1,F}\}$), four unique uncertainties (i.e., \textit{UU(BM)}=$\{un_1, un_2, un_3, un_4\}$), and two uncertainty spaces (i.e., \textit{USP(BM)}=$\{usp_1, usp_2\}$). 
}

\subsection{Test cases generation with UncerTest}\label{sec:testCaseUncerTest}

\textit{UncerTest}~\cite{zhang2019uncertainty} is a model-based test case generation approach for testing CPSs under uncertainty. Specifically, \textit{UncerTest} takes BMs (Section~\ref{sec:beliefModel}) as input and generates a set of executable test cases with uncertainty sources (i.e., indeterminacy sources) seeded in test environments.

\subsubsection{Uncertainty measurements 
}\label{sec:uncertaintyMeasurement}

Uncertainty can be measured differently. 
For instance, \textit{Probability Theory} is commonly applied to measure uncertainty as a frequency.
However, its application requires a large amount of data collected from a long-run experiment to make the estimated measurement of uncertainty (i.e., \textit{probability}) ``close enough to the long-run frequency''. 
However, such a large amount of data is usually unavailable at the startup phase of the test design in the context of MBT~\cite{liu2012there}. 
Therefore, \textit{UncerTest} opts for \textit{Uncertainty Theory} for handling uncertainty measurement when sufficient data does not exist to estimate the \textit{probability}. 
This theory measures uncertainty as a \textit{belief degree} from the subjective perspective of belief agents (engineers in our context). 

\textit{Probability Theory} measures uncertainty as \textit{probability}, while 
\textit{Uncertainty Theory} measures uncertainty as \textit{Uncertain Measure}
\textit{Uncertain Measure} is represented as the $\mathcal{M}$ symbol and respects the three axioms below~\cite{liu2007uncertainty}: 
\begin{compactitem}
\item \textbf{Normality}: $\mathcal{M}\revision{\{}\Gamma\revision{\}}=1$, ($\Gamma$ is the universal set).

\item \textbf{Duality}: $ \mathcal{M}\left\{\Lambda\right\}+\mathcal{M}\{{\Lambda}^c\}=1$, where $\Lambda$ denotes a particular event, whereas ${\Lambda}^c$ denotes all the events in the universal set excluding $\Lambda$.

\item \textbf{Subadditivity}: $\mathcal{M}\{\bigcup^{\infty}_{i=1}{{\Lambda}_i}\}<\sum^{\infty}_{i=1}{\mathcal{M}\left\{{\Lambda}_i\right\}}$, (every \textit{countable} sequence of events ${\Lambda}_1,{\Lambda}_2,\dots$).
\end{compactitem}

\noindent\textit{Uncertainty Space} is defined as triplet ($\Gamma, \mathcal{L}, \mathcal{M}$), where $\Gamma$ is the universal set, $\mathcal{L}$ is a $\sigma$-algebra~\cite{walter1987real} over $\Gamma$. 

\textbf{Theorem}: Let ($\Gamma_k, \mathcal{L}_k, \mathcal{M}_k$) be uncertainty spaces and ${\Lambda}_k\in{\mathcal{L}}_k$, for $k=1, 2, \dots n$. Then ${\Lambda}_{1}, {\Lambda}_{2}, \dots {\Lambda}_n$ are always independent of each other if they are from different uncertainty spaces.

\revision{
As shown in Figure~\ref{fig:example}, with \textit{Probability Theory}, 
uncertainty $un_2$ is calculated with its occurrence frequency. For instance, $un_2$ occurs 98\% of the time, hence \textit{UM}$(un_2)$= \textit{Pr}$(un_2)$=$0.98$.  
With \textit{Uncertainty Theory}, we consider $un_1$ and $un_2$ from uncertainty space $usp_1$, and $\mathcal{L} =\{\emptyset,\{un_1\}, \{un_2\}, \{un_1, un_2\}\}$.
The measurement is calculated with \textit{belief degree} of domain experts (e.g., when there is insufficient historical data to calculate the frequency),
for instance, \textit{UM}$(un_2) = \mathcal{M}\{un_2\} = 0.98$.
The uncertainties from different uncertainty spaces (e.g., $un_2$ from $usp_1$ and $un_3$ from $usp_2$) are independent of each other.}


\subsubsection{Enabling Indeterminacy Sources}\label{sec:enablingIndeterminacy}
In \textit{UncerTum}'s \textit{UUP}, \textit{IndeterminacySource} is for constructing known sources of uncertainty. We extended it to trigger occurrences of specified uncertainties during test execution by enabling indeterminacy sources in generated executable test cases. 
\revision{For instance, it might increase the chance of enabling the occurrence of uncertainty $un_3$\textit{(S1, T2, S1)} (Figure~\ref{fig:example}) by employing a motion sensor jammer (an indeterminacy source \textit{IndS}$_1$) when introducing an intrusion (see \textit{T2}$^{S1,S1}$).}
In addition, indeterminacy sources are parts of assertions for uncertainty occurrences, e.g., checking whether an uncertainty occurs with enabled indeterminacy sources. More details are provided in Section~\ref{sec:testResults}.

\subsubsection{Generating Uncertainty-aware Test Cases}\label{sec:testCaseUncertainty}

\revision{The need for uncertainty-wise testing emerged as part of our cooperation with industrial partners, who observed that during the operation of their systems, they observe uncertain behaviors (e.g., a device showing wrong measurements for a short period, raising false alarms). As a result, they wanted to test their system to find uncertainties. To this end, in our previous work, we proposed  \textit{UncerTest}~\cite{zhang2019uncertainty}} consisting of two test case generation strategies, \textit{All Simple Belief Paths} (\textit{ASiBP}) and \textit{All Specified Length Belief Paths} (\textit{ASlBP}), for generating test cases based on BMs. The two strategies are distinguished by whether \textit{round trips} \revision{(i.e., paths containing circles)} \textit{are allowed} from the practical perspective.

\begin{table}[!t]
\small
\caption{Characterization of test cases being prioritized}
\label{table:testCaseAttribute}
\centering
\resizebox{0.99\linewidth}{!}{
\input{tables/tabTestCaseAttribute}
}
\end{table}

Since BMs inputted to \textit{UncerTest} contain uncertainty information, each generated test case naturally has uncertainty characterized by the first five attributes in Table~\ref{table:testCaseAttribute}. For each test case, the values of the first three attributes can be easily obtained by counting the number of transitions and (unique) uncertainties along the test case's path.  
An \usp groups uncertainties originating from the same state with the state transition representing its potential outcomes.
\usp{\textit{(t)}} counts a number of such \usp covered in the path of the test.
\um{\textit{(t)}} represents its uncertainty measurement derived based on \textit{Uncertainty Theory} (Section~\ref{sec:uncertaintyMeasurement}).
\revision{For instance, based on the BM shown in Figure~\ref{fig:example}, test $t_1$ traversing \textit{I}$\rightarrow$\textit{T0}$\rightarrow$\textit{S0}$\rightarrow$\textit{T1}$\rightarrow$\textit{S1}$\rightarrow$\textit{T2}$\rightarrow$\textit{S2}$\rightarrow$\textit{T3}$\rightarrow$\textit{S1}$\rightarrow$\textit{T2}$\rightarrow$\textit{S2}$\rightarrow$\textit{T2f}$\rightarrow$\textit{F} could be derived with the \textit{ASlBP} strategy, which contains \textit{round trips} (e.g., \textit{S1}$\rightarrow$\textit{T2}$\rightarrow$\textit{S2}$\rightarrow$\textit{T3}$\rightarrow$\textit{S1}). Test $t_1$ covers five transitions (i.e., \textit{TR}$(t_1)=\{$\textit{T0}$^{I, S0},$\textit{T1}$^{S0,S1}, $\textit{T2}$^{S1,S2}, $\textit{T3}$^{S2,S1}, $\textit{T2f}$^{S2,F}\}$) out of the eight, three uncertainties (i.e., \textit{Us}$(t_1)$=$\{un_2, un_4, un_4\}$), two unique uncertainties (i.e., \textit{UU}$(t_1)$=$\{un_2, un_4\}$) out of the four, and two uncertainty spaces (i.e., \textit{USP}$(t_1)$=$\{usp_1,usp_2\}$). By applying \textit{Probability Theory}, the measurement is 0.96 (i.e., \textit{UM}$(t_1)$=\textit{Pr}$(t_1)$= $0.98 \times 0.99 \times 0.99$ = $ 0.96$) while the measurement is 0.98 by applying \textit{Uncertainty Theory} (i.e., \textit{UM}$(t_1) $=$ \mathcal{M}\{t_1\} $=$ 0.98 \wedge 0.99 \wedge 0.99$=$0.98$).
} 

\textit{UncerTest} also converts indeterminacy sources captured along the sequence of model elements of the BM (e.g., state, transition) as part of a generated executable test case. 
\textit{UncerTest} is equipped with four search-based test case minimization strategies, which minimize the number of test cases to execute and maximize the number of uncertainties.
\revision{The minimization of \textit{UncerTest} aims at reducing test cases from tests generated automatically based on specified preference (e.g., high uncertain measure).
However, such a minimization does not consider the practical constraints of executing all tests even after being minimized, e.g., time budget. As a result, we observed that it was too costly to execute even the minimized tests while working with our industrial partners. Therefore, there was an emergent need to execute those test cases as soon as possible, which were likely to lead to observing uncertainties. To cater to such a need, we propose an uncertainty-aware prioritization specific to handle uncertainty-aware test case execution by considering time-related execution costs.}

\subsection{Uncertainty-aware Test Verdict}\label{sec:testResults}
\textit{UncerTest} also has a set of uncertainty-aware test verdicts for assessing the occurrence of uncertainties together with the occurrence of indeterminacy sources (Section~\ref{sec:enablingIndeterminacy}). \revision{For instance, executing $t_1$ as shown in Figure~\ref{fig:example}, when the SUT firstly arrives at state \textit{S1} (i.e., \textit{I}$\rightarrow$\textit{T0}$\rightarrow$\textit{S0}$\rightarrow$\textit{T1}$\rightarrow$\textit{S1}), after an intrusion is introduced to the SUT (i.e., \textit{T2}), a specified uncertainty (i.e., $un_4$) is considered as \textit{occurred} (i.e., \textit{KnOcurred}) if the state of the SUT transits to \textit{S2} while a specified uncertainty (i.e., $un_4$) is considered \textit{not occurred} (i.e., \textit{KnNotOcurred}) if the state of the SUT transits to \textit{S1}, an alternative specified uncertainty in its uncertainty space. Note that if the SUT transits to neither \textit{S1} or \textit{S2}, we consider an unknown uncertainty \textit{occurred}, i.e., \textit{UkOccurred}. The occurrence of indeterminacy sources depends on whether the uncertainty is associated with an indeterminacy source (e.g., \textit{IndS}$_1$ for $un_3$) and whether the indeterminacy source is enabled or not when the uncertainty occurs. Note that the occurrence of an uncertainty might be nondeterminate, e.g., due to a lack of knowledge of all associated indeterminacy sources; hence, execution results for the same uncertainty might be different in multiple executions.} Evaluation results are part of test execution results collected during test execution. 
\textit{UncerTest}'s uncertainty-aware verdicts are:
\begin{itemize}
    \item \textit{KnOccurred-With-InS}: a specified uncertainty occurred under the occurrence of one or more specified indeterminacy sources.
    \item \textit{KnOccurred-Without-InS}: a specified uncertainty occurred without having any specified indeterminacy source occurred.
    \item \textit{KnNotOccurred-With-InS}: a specified uncertainty did not occur even though one or more specified indeterminacy sources occurred.
    \item \textit{KnNotOccurred-WithoutInS}: a specified uncertainty did not occur under the non-occurrence of any specified indeterminacy source.
    \item \textit{KnOccurred-UkInS}: a specified uncertainty occurred, and its indeterminacy source(s) is unspecified (i.e., unknown).
    \item \textit{KnNotOccurred-UkInS}: a specified uncertainty did not occur, and its indeterminacy source is unspecified (i.e., unknown).
    \item \textit{UkOccurred}: a specified uncertainty did not occur, and none of 
    the existing model elements (i.e., states and transitions in the \textit{BSM})
    could match the occurrence, i.e., an unknown uncertainty occurred. 
\end{itemize}

\revision{Compared to traditional test verdicts (e.g., \textit{Fail}, \textit{Pass}), uncertainty-aware verdicts are defined for identifying \textit{unknowns}~\cite{zhang2019uncertainty}, i.e., observing an uncertainty occurrence that is previously unknown or a known uncertainty occurred with previously unknown indeterminacy sources. Such \textit{unknowns} might refer to existing \textit{defects}. For instance, an uncertainty that the SUT has not properly handled because its source (i.e., \textit{indeterminacy source}) was previously unknown, which might relate to a flaw in the design of the SUT, e.g., due to a lack of knowledge about the complexity of the system itself and the complicated environment in operation. Whether to identify an observed \textit{unknown} as a \textit{defect} or not requires determinations from practitioners who should have the correct mindset of considering that recognizing uncertainties is essential for reducing the risk of defects. Moreover, it is interesting to study whether uncertainties relate to existing bugs or not requires further investigation by practitioners. Consequently, studying such correlations to guide practitioners in better identifying and locating bugs requires conducting an entirely new empirical study, which is not the scope of this paper. We would also like to mention that these uncertainty-aware verdicts are especially important for CPSs, with some of their behaviors that can only be known after deployment and in operation. }



\section{Approach}\label{sec:problems}

\begin{figure*}[!t]
\centering
\includegraphics[width=0.8\textwidth]{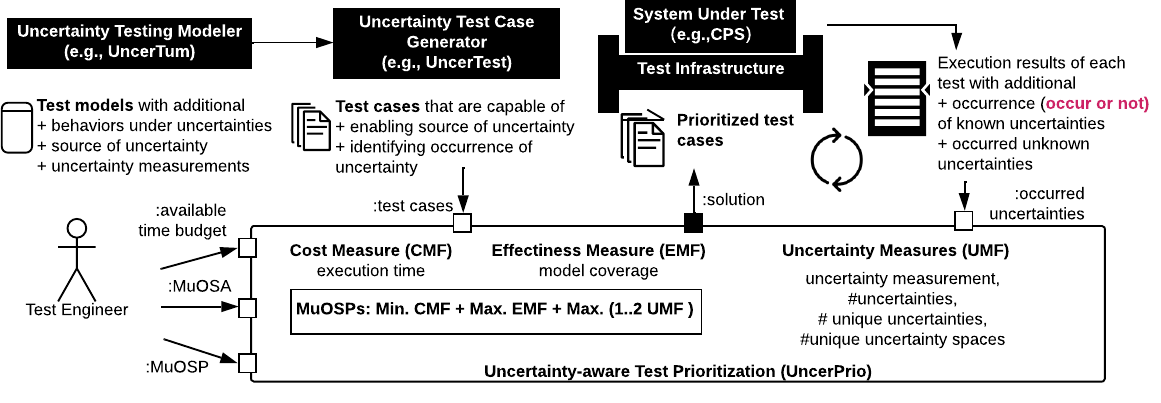}
\begin{spacing}{-0.5}
\end{spacing}
\begin{threeparttable}
\begin{tablenotes}
\scriptsize \item[*] MuOSP: Multi-Objective Search Problem; MuOSA: Multi-Objective Search Algorithm
\end{tablenotes}
\end{threeparttable}
\caption{Overview of \approach}
\label{fig:overview}
\end{figure*}

Figure~\ref{fig:overview} presents an overview of \approach, together with \uncertum and \uncertest, severing as an integrated platform for testing CPSs under uncertainties.

First, uncertainties can be captured as part of BMs with \uncertum for specifying uncertainty-related information, such as potential uncertain behavior resulting from known indeterminacy sources and uncertainty measurements (subjectively indicating belief agents' belief degrees).
Then, \uncertest generates tests and minimizes them with a strategy \revision{specified by users based on their preferences}. Generated test cases are embedded with triggers of related indeterminacy sources and verdicts for observing uncertainty occurrences. Third, the execution results of such tests allow \approach to collect objective uncertainty measurements as frequencies of the occurrence of subjective uncertainties specified in the BMs and unknown uncertainties (i.e., those previously unspecified in the BMs).

To solve the test optimization problem with \MOSA{s}, we formulate the \uncertaintyaware test case prioritization problem as search problems (Section~\ref{sec:problemFormulation}),
define the time cost measure (\textit{CMF}), effectiveness measure (\textit{EMF}), and uncertainty measures (\textit{UMF}) (Section~\ref{sec:costEffectivenessMeasure}), and propose the fitness function to minimize \textit{CMF}, maximize \textit{EMF} and maximize \textit{UMF} (Section~\ref{sec:fitnessFunction}).

%

\subsection{Problem Formulation}\label{sec:problemFormulation}
Let \textit{T} = $\{$\textit{t${}_{i}$}{\textbar} 0 $<$ \textit{i} $\leq$ \textit{nt}$\}$ be a set of test cases derived from a BM with \textit{UncerTest} (Section~\ref{sec:background}), which need to be prioritized. As discussed in Section~\ref{sec:background}, each test case $t_{i}$ in \textit{T} is characterized by the six attributes (Table~\ref{table:testCaseAttribute}). Thus, a prioritized solution within a specific time budget can be defined as below:
\begin{equation*}
T_{prio} = \left\{t^{'}_{j}, tb\right| 0 < j \le mt, \sum^{mt}_{1} ET\left(t^{'}_{j}\right) \le tb \times ET(T)\} \subseteq T
\end{equation*}
where $T_{prio}$ is a subset of \textit{T}, \textit{j} is the position index \revision{(its order to execute)} of the prioritized test cases, \textit{mt} is the total number of test cases, \textit{ET(t)} is the execution time of test case \textit{t}, \textit{ET(T)} is the total execution time of all the test cases \textit{T}, and \textit{tb} refers a time budget defined as the percentage of the total execution time \textit{ET(T)}. 
For example, \textit{tb = 40\%} means that the execution time of the prioritized test cases cannot be more than 40\% of the execution time of all the test cases.

We define \textit{PI(j)} to measure the impact of position \textit{j} of test case \textit{${t}^{'}_{j}$} in $T_{prio}$ based on its cost-effectiveness. The impact can be calculated as: $PI\left(j\right)=\frac{mt-j+1}{mt}, 0<j\le mt$. The overall idea is to guide the search to prioritize test cases with high cost-effectiveness in earlier positions to be executed.

The search space of the test case prioritization problem is the set of permutations of all the test cases from \textit{T} or a subset of it, as represented below:
\begin{equation*}
PS = \left\{{ps}_1, ps_2, ps_3 \dots {ps}_{nps}\right\}
\end{equation*}
where ${ps}_{i}$ is one potential solution $T_{prio}$, and $nps$ is the total number of potential solutions.
Each solution ${ps}_i$ can be characterized with the following list of measures:
\begin{itemize}
\item \textit{CostMeasure} $=\left\{{cm}_1,{cm}_2, \dots , {cm}_{ncm}\right\}$, a set of cost measures;
\item \textit{EffectMeasure} $=\left\{{em}_1,{em}_2, \dots , {em}_{nem}\right\}$, a set of effectiveness measures; and 
\item \textit{UncerMeasure} $=\{{um}_1,\ {um}_2, \dots , {um}_{num}\}$, a set of \uncertaintyaware measures.
\end{itemize}

\noindent\textbf{Uncertainty-aware Test Case Prioritization Problem:} With given time budget \textit{tb}, find a set of Pareto optimal solutions ${PS}_p=\left\{{ps}_k\mathrel{\left|\vphantom{{ps}_k k\ge 1}\right.\kern-\nulldelimiterspace}k\ge 1\right\}\subseteq PS\ $that aim at:
\begin{equation*}
min\left({CMF}_1\left(ps\right),\dots ,{CMF}_{ncm}\left(ps\right)\right), 
\end{equation*}
\begin{equation*} 
max\left({EMF}_1\left(ps\right),\dots ,{EMF}_{nem}\left(ps\right)\right), and
\end{equation*}
\begin{equation*}
max({UMF}_1\left(ps\right),\dots ,{UMF}_{num}(ps)),
\end{equation*}
where 1) $ncm \geq 1$ and ${CMF}_{i}(ps)$ are for calculating values of the cost measure ${cm}_{i}$ for solution $ps$; 2) $nem \geq 1$ and ${EMF}_{i}(ps)$ are for calculating values of the effectiveness measure ${em}_{i}$ for $ps$; 3) $num \geq 1$ and ${UMF}_{i}(ps)$ are for calculating values of the uncertainty measure ${um}_{i}$) for $ps$.
${PS}_{p}$ is a set of nondominated solutions and there does not exist a solution $ps \in PS$ that dominates ${ps}_{k} \in {PS}_{p}$, i.e., $\nexists ps \in PS - {PS}_{p}$ that satisfies 
    1) \revision{$\forall CMF_{i} \in \{CMF_1,...,CMF_{ncm}\}$}, 
${CMF}_{i}(ps) \le {CMF}_{i}({ps}_{k})$, and
    2) \revision{$\forall EMF_{i} \in \{EMF_1,...,EMF_{nem}\}$}, ${EMF}_{i}(ps) \geq {EMF}_{i}({ps}_{k})$, and
    3) \revision{$\forall UMF_{i} \in \{UMF_1,...,UMF_{num}\}$}, ${UMF}_{i}(ps) \geq {UMF}_{i}({ps}_{k})$, and
    4) \revision{$\exists CMF_{i} \in \{CMF_1,...,CMF_{ncm}\}$}, ${CMF}_{i}(ps) < {CMF}_{i}({ps}_{k})$, or 
    \revision{$\exists EMF_{i} \in \{EMF_1,...,EMF_{nem}\}$}, ${EMF}_{i}(ps) > {EMF}_{i}({ps}_{k})$, or 
    \revision{$\exists UMF_{i} \in \{UMF_1,...,UMF_{num}\}$}, ${UMF}_{i}(ps) > {UMF}_{i}({ps}_{k})$.

\subsection{Definitions of Measures}\label{sec:costEffectivenessMeasure}
We introduce six measures to formulate the uncertainty-aware test prioritization problem: one cost, one effectiveness and four uncertainty measures. 
Note that the value ranges of all the measures are normalized between 0 and 1.

\subsubsection{Cost Measure}\label{sec:costMeasure}
\textbf{C1.	Percentage of Execution Time (PET)}
$PET\left(T_{prio}\right)$ measures time cost spent by tests as a
percentage of the total execution time of all the test cases (i.e., $\frac{ET(t^{'}_j)}{ET_{total}}$ for a test at $j$th position),
which can be calculated as:
\begin{equation}
PET\left(T_{prio}\right)=\frac{\sum^{mt}_{j=1} ET(t^{'}_j) \times PI(j)}{{ET}_{total}}, 0<j\le mt
\end{equation}
We aim to minimize \textit{PET}, i.e., \textit{PET}$\downarrow$, spent by the prioritized test cases; thus, a smaller value indicates a better result.

\subsubsection{Effectiveness Measures}\label{sec:effectivenessMeasure}
\textbf{E1.	Percentage of Transition Coverage (PTR)}
$PTR\left(T_{prio}\right)$ measures the percentage of the total number of unique transitions covered by the prioritized test cases. We first define the set of unique transitions covered by $T^{'}_{j}$ as: ${TR}^{'}\left(T^{'}_{j}\right)=\bigcup^{j}_{k=1}{TR(t^{'}_{k})}$, and the size of ${TR}^{'}\left(T^{'}_{j}\right)$ as ${ntr}^{'}_{{T}^{'}_{j}}$. The relative complement of ${TR}^{'}\left(T^{'}_{j-1}\right)$ in ${TR}^{'}\left(T^{'}_{j}\right)$ is then defined as: 
\begin{equation}
\Delta{TR}^{'}\left(T^{'}_{j}\right)=\left\{\begin{array}{ll} {TR}^{'}\left(T^{'}_{j}\right)\setminus{TR}^{'}\left(T^{'}_{j-1}\right), & j>1\\ {TR}^{'}\left(T^{'}_{j}\right), & j=1 \end{array}\right.
\end{equation}
where $\Delta {ntr}^{'}_{T^{'}_{j},1}$ is the size of ${\Delta TR}^{'}\left(T^{'}_{j}\right)$. Based on these definitions, we can therefore calculate $PTR\left(T_{prio}\right)$ as:
\begin{equation}
PTR\left(T_{prio}\right) = \frac{\sum^{mt}_{j=1}{\Delta ntr^{'}_{T^{'}_{j},1}\times PI(j)}}{ntr}, 0<j\le mt
\end{equation}
where $ntr$ is the total number of transitions covered by \textit{BSM} used for deriving the test cases. 
\revision{A higher \textit{PTR} value indicates, in test execution, to prioritize tests that include more unique transitions, i.e., covering more SUT behaviors. Hence, we} aim to maximize \textit{PTR} (denoted as \textit{PTR}$\uparrow$) covered by prioritized test cases with a higher value indicating a better result.

\subsubsection{
Uncertainty measures}\label{sec:uncertaintyMeasure}
\revision{To consider \textit{uncertainty} from various aspects in testing, we define four uncertainty-aware objectives measuring uncertainty from four uncertainty-related characteristics (see Table~\ref{table:testCaseAttribute}), i.e., the measurement of its occurrence (i.e., \textit{UM}), the quantity of uncertainties being tested (i.e., \textit{Us}), the coverage of uncertainties (i.e., \textit{UU}), and the coverage of kinds of uncertainties (i.e., \textit{USP}) in the context of MBT.}


%
%


\textbf{U1.	
Average Uncertainty Measure (AUM)}
$AUM\left(T_{prio}\right)$ measures the average of 
uncertainty measurement of tests
in the prioritized subset of test cases:
\begin{equation}
AUM\left(T_{prio}\right)=\frac{\sum^{mt}_{j=1}{UM(t^{'}_{j})}\times PI(j)}{mt}, 0<j\le mt
\end{equation}
where $mt$ is the total number of test cases in $T_{prio}$ and $UM(t^{'}_{j})$ is the 
uncertainty measurement of $t^{'}_{j}$.
As \textit{Uncertainty Theory} used in \textit{UncerTest} (see Table~\ref{table:testCaseAttribute}), 
$UM(t^{'}_{j})$
measures a modeler's confidence that the test $t^{'}_{j}$ will pass (i.e., the uncertainties in the test will occur as specified).
\revision{A higher value indicates, in the test execution, to prioritize tests that are more likely to pass by including uncertainties with higher measurements, then we} aim to maximize \textit{AUM} (denoted as \textit{AUM}$\uparrow$); thus, a higher value of it indicates a better solution. 

\noindent\textbf{U2.	
Average Normalized Number of Predefined Uncertainties Covered (ANU)}
$ANU\left(T_{prio}\right)$ measures the average normalized number of 
uncertainties that exist in the prioritized test cases:
\begin{equation}
ANU\left(T_{prio}\right)=\frac{\sum^{mt}_{j=1}{nor\left({nu}_{t^{'}_{j}}\right)\times PI(j)}}{mt},\ 0<j\le mt
\end{equation}
where ${nu}_{t^{'}_{j}}$ is the number of 
uncertainties specified in test case $t^{'}_{j}$ which can be normalized as $nor\left(x\right)=\frac{x}{x+1}$, and $mt$ is the total number of test cases in $T_{prio}$. 
\revision{A higher value indicates, in the test execution, to prioritize tests including more uncertainties to increase the chance to test the SUT in the presence of uncertainties, then we} aim to maximize \textit{ANU} covered by the prioritized test cases (denoted as \textit{ANU}$\uparrow$); thus, a higher value indicates a better solution. 

\noindent\textbf{U3.	
Percentage of Unique Predefined Uncertainties Covered (PUU)}
$PUU\left(T_{prio}\right)$ measures the percentage of the total number of unique uncertainties covered by prioritized test cases. We define the set of unique uncertainties covered by $T^{'}_{j}$ as: ${UU}^{'}(T^{'}_{j})=\bigcup^{j}_{k=1}{UU(t^{'}_{k})}$, and its size is ${nuu^{'}}_{T^{'}_{j}}$. The relative complement of $UU^{'}\left(T^{'}_{j-1}\right)$ in $UU^{'}\left(T^{'}_{j}\right)$ is then defined as: 
\begin{equation}
\Delta UU^{'}\left(T^{'}_{j}\right) = \left\{\begin{array}{ll} UU^{'}\left(T^{'}_{j}\right){\setminus}UU^{'}\left(T^{'}_{j-1}\right), & j>1\\ UU^{'}\left(T^{'}_{j}\right), & 0<j\le 1 \end{array}\right.
\end{equation}
where $\Delta {nuu}^{'}_{T^{'}_{j},1}$ is the size of $\Delta UU^{'}\left(T^{'}_{j}\right)$. Based on these definitions, we can therefore calculate $PUU\left(T_{prio}\right)$ with:
\begin{equation}
PUU(T_{prio})=\frac{\sum^{mt}_{j=1}{\Delta {nuu}^{'}_{T^{'}_{j,1}}\times PI(j)}}{nuu}, 0<j\le mt
\end{equation}
where $nuu$ is the total number of uncertainties covered by the \textit{BSM}. 
\revision{A higher value indicates, in the test execution, prioritize tests that cover more unique uncertainties, then}
we aim to maximize \textit{PUU} covered by the prioritized test cases (denoted as \textit{PUU}$\uparrow$); thus, a higher value indicates a better solution. 

\noindent\textbf{U4.	
Percentage of Uncertainty Space Covered (PUS)}
$PUS(t^{'}_{j})$ is the percentage of the total set of uncertainty spaces of the \textit{BSM} covered by test case $t^{'}_{j}$, which can be calculated as:
\begin{equation}
PUS\left(t^{'}_{j}\right)=\frac{{musp}_{t^{'}_{j}}}{nusp}\times 100\%
\end{equation}
where $musp$ is the number of predefined uncertainty spaces covered by test case $t^{'}_{j}$ and $nusp$ is the total number of predefined uncertainty spaces in the \textit{BSM}.
$PUS\left(T_{prio}\right)$ measures the percentage of uncertainty spaces covered by the prioritized test cases, which can be calculated as:
\begin{equation}
PUS\left(T_{prio}\right)=\frac{\sum^{mt}_{j=1}{PUS(t^{'}_{j})}\times PI(j)}{mt}, 0<j\le mt
\end{equation}
where $mt$ is the total number of test cases in $T_{prio}$. 
\textit{PUS} is a measure defined based on \textit{Uncertainty Space} in Uncertainty Theory (Section~\ref{sec:background}). 
\revision{A higher value indicates, in the test execution, prioritize tests covering more diverse kinds of uncertainties (i.e., more unique uncertainty spaces), then}
we aim to maximize \textit{PUS} achieved by the prioritized test cases (\textit{PUS}$\uparrow$); thus a higher value indicates a better solution.

\subsection{Fitness Functions}\label{sec:fitnessFunction}

\revision{We reformulate the test prioritization problems as multi-objective search problems, and each is considered a strategy to prioritize uncertainty-aware tests for test execution.}
Considering the cost-effectiveness and 
being \uncertaintyaware \revision{(i.e., to trade-off between cost, model coverage, and uncertainty-aware effectiveness),} we define 10 prioritization problems and formalize them as 10 fitness functions, as shown in Table~\ref{table:tabFitnessFunction}\revision{, aiming to identify the best strategy for uncertainty-aware test prioritization}.
All of the prioritization problems were formed with the cost measure (\textit{PET}), the effectiveness measure (\textit{PTR}), and one or two uncertainty measures of the four. 
\revision{The rationale is to consider potential correlations between any two uncertainty measures and study their effectiveness in solving uncertainty-aware test prioritization.}
For instance, 
\revision{\probFifth} is to prioritize test cases \revision{combined with two uncertainty measures} that take less time to execute, cover more transitions, have a higher 
uncertainty measurement (i.e., a higher \textit{confidence} that uncertainties will occur as specified)\revision{, and cover more kinds of uncertainties (i.e., more uncertainty spaces)}
in the front of the test execution sequence.  

To be time-aware, we also define 10 time budgets (denoted as TB10{\dots}TB100), formulated as constraints in search, i.e., the execution time of $T_{prio}$ should be within a given time budget, as shown as: 
\begin{equation}
    ET \leq tb \% \times ET_{total}, tb=10, 20\dots100
\end{equation}
where $ET$ is the execution time of $T_{prio}$ and $ET_{total}$ is the total execution time of all the test cases to prioritize $T$.
\revision{Time constraints restrict how many tests can be executed in the prioritized solution $T_{prio}$.}
\begin{table}[!t]
\small
\caption{Prioritization Problems and Fitness Functions}
\label{table:tabFitnessFunction}
\centering
\resizebox{0.99\linewidth}{!}{
\input{tables/tabFitnessFunction}
}
\end{table}
%

\section{Evaluation}\label{sec:evaluation}
Section~\ref{sec:experimentDesign} presents the experiment design, followed by results\revision{, a recommendation to apply \approach based on the results} and threats to validity in Sections \revision{\ref{sec:experimentResult}} and\revision{~\ref{sec:threatValidity} }.

\subsection{Experiment Design}\label{sec:experimentDesign}
The experiment design is summarized in Table~\ref{table:tabExperimentDesign}, which shows that our empirical study has 2500 settings (= 10 × 10 × 5 × 5), formed by considering the 10 prioritization problems with the 10 time budgets solved by 5 algorithms (RS and 4 \MOSA{s}) for 5 use cases of 2 subject systems.

In Section~\ref{MOSAs}, we present the employed \MOSA{s}. In Section~\ref{sec:industrialSystem}, we present the subject systems. In Sections~\ref{sec:researchQuestion} and ~\ref{metrics}, we present the research questions (RQs) and evaluation metrics, followed by the statistical tests (Section~\ref{sec:statisticalTest}).

\begin{table*}[!t]
\caption{Experiment Design}
\label{table:tabExperimentDesign}
\centering
\resizebox{0.95\textwidth}{!}{
\input{tables/expDesign}
}
\end{table*}

\subsubsection{Employed \MOSA{s}}\label{MOSAs}
The employed \MOSA{s} are NSGA-II~\cite{deb2002fast}, MOCell~\cite{nebro2009mocell}, SPEA2~\cite{zitzler2001spea2}, and CellDE~\cite{durillo2008solving}, which are implemented in jMetal~\cite{jMetal} and have been applied for addressing various software engineering optimization problems (e.g., requirements engineering optimization problems~\cite{li2017zen, yue2014applying}, product configuration~\cite{lu2016nonconformity} and testing~\cite{wang2016enhancing, wang2014multi}). 
RS is used as the comparison baseline to justify the use of \MOSA{s}.

For all of the algorithms,
we used their default parameter settings in jMetal, which are also provided in Table~\ref{table:tabParameterSetting} for reference. The stopping criterion is the 25000 times fitness evaluations.
Considering the inherent randomization of search algorithms, we run RS and each \MOSA 100 times.

\subsubsection{Subject Systems}\label{sec:industrialSystem}

The Empirical study of our prioritization problem is conducted with two industrial case studies. The first system is GeoSports (\textit{GS}), which attaches devices to Bandy\footnote{Bandy is a variation of ice hockey often played in Northern Europe.} players for recording measurements (e.g., heartbeat, speed, location) and deliver them during a game via a receiver station to a runtime monitoring system used by coaches. To test GS in a lab setting without real players, a test infrastructure (including hardware and software) was developed to execute test cases~\cite{zhang2019uncertainty}. Executing test cases on this test infrastructure is both time-wise and monetary expensive; hence, available test suites need to be optimized.

Another subject system is Automated Warehouse (\textit{AW}) equipped with various handling facilities (e.g., cranes, conveyors, sorting systems, picking systems, rolling tables, lifts, and intermediate storage). A cloud-based supervision system interacts with these physical units and network equipment. \textit{AW} implements several key scenarios such as introducing many pallets to the warehouse and transferring items with stacker cranes. To test such scenarios, relevant simulators and emulators have been developed. Further details on the subject systems can be consulted in~\cite{U-Test}.

Test cases generated and minimized with \textit{UncerTest} for \textit{GS} and \textit{AW} are used as the dataset for evaluating \approach, and \revision{the description of BMs used for test generation can be found in~\cite{U-Test}}. The descriptive statistics of the test cases collected from the five use cases of the two systems are provided in Table~\ref{table:tabStatistics}. Column ``\# of Test Cases" shows the number of test cases to prioritize. For example, for \textit{AW1}, the total number of executable test cases to be prioritized is 420, which requires 7924 seconds to execute.

\begin{table}[!t]
\caption{Characteristics of Employed Test Case Datasets}
\label{table:tabStatistics}
\centering
\resizebox{0.46\textwidth}{!}{
\input{tables/tabStatistics}
}
\end{table}

\subsubsection{Research Questions}
\label{sec:researchQuestion}
In this empirical study, we aim to answer four RQs:

\begin{compactitem}[-]
\item \textbf{RQ1:}
How do the selected \MOSA{s} perform, when compared to RS, in terms of solving the 10 \uncertaintyaware test case prioritization problems constrained with the 10 time budgets? 
\item \textbf{RQ2:} For each use case and each time budget, which \MOSA is the best in solving each prioritization problem? 
\item \textbf{RQ3:} For each use case, how do the time budgets impact MOSAs' performance in solving each problem? 
\item \textbf{RQ4:} For each SUT with a given time budget, which prioritization problem is most efficient in observing uncertainties?
\end{compactitem}
RQ1 is simply for the sanity check such that the use of \MOSA{s} can be justified. RQ2 is defined for comparing the performance of the \MOSA{s} in solving the optimization problems.  
RQ3 is designed to investigate the impact of the various time budgets on the results.
RQ4 is for studying the efficiency of each test case prioritization problem in terms of observing uncertainties.

\subsubsection{Evaluation Metrics}\label{metrics}
To assess the performance of the \MOSA{s}, we apply Hypervolume (\textit{HV}) \revision{and Inverted Generational Distance (\textit{IGD}) indicators to compare the performance of search algorithms according to the guideline in~\cite{ali2020quality}.} 
\textit{HV} computes the volume in the objective space that is covered by a non-dominated set of solutions (e.g., Pareto front), by considering both convergence and diversity. 
\revision{\textit{IGD} measures the distance between the optimal non-dominated set of solutions and the nearest solutions in the computed non-dominated set of solutions.}


Based on the \uncertaintyaware test verdict (Section~\ref{sec:testResults}), for each test, we define a number of observed uncertainties \textit{NOU}$_t$ as a basic evaluation metric. Based on \textit{\nou}, to enable comparison of solutions across the time budgets and problems, we define Average Number of Observed Uncertainties (\anou).
\anou$\left(T_{prio}\right)$ measures the average number of observed uncertainties by the prioritized test cases, as formally defined below:
\begin{equation}
ANOU\left(T_{prio}\right)=\frac{\sum^{mt}_{j=1}{nou_{t^{'}_{j}}\times PI(j)}}{mt}, 0<j\le mt
\end{equation}
where ${nou}_{t^{'}_{j}}$ is the number of uncertainties observed in executing test case $t^{'}_{j}$, and $mt$ is the total number of test cases in $T_{prio}$. 
This metric is an objective uncertainty measurement and is used to assess the efficiency of solutions (i.e., prioritized test cases) achieved by the prioritization problems we defined and the \MOSA{s} we employed.

\begin{table}[!t]
\caption{Parameter Settings of the \MOSA{s}}
\label{table:tabParameterSetting}
\centering
\resizebox{0.95\linewidth}{!}{
\input{tables/tabParameterSetting_new}
}
\end{table}

\subsubsection{Statistical Tests}\label{sec:statisticalTest}
Table~\ref{table:tabExperimentDesign} represents the statistical tests we applied for answering the RQs.
%
Based on the guidelines in~\cite{dodge2008concise, arcuri2011practical}, we conduct the comparative analysis with the Kruskal–Wallis test, Mann-Whitney U test, and the Vargha and Delaney statistics.
Our study compares two or more groups (e.g., algorithms or problems) with the Kruskal–Wallis test.
If comparison results show that at least one group originates from a different distribution compared to the others (i.e., $p <$ 0.05), we apply the Mann-Whitney U test to perform pair-wise comparisons with a significance level of 5\%.
Besides, we used the Vargha and Delaney statistics to calculate \Atwelve, a non-parametric effect size measure, to demonstrate which group of a pair gives better results. 
\revision{
To deal with aggregated error probabilities arising due to multiple comparisons, we further applied the Holm–Bonferroni method~\cite{holm1979simple} as a post-hoc analysis to control the family-wise error rate at level $\alpha$ (i.e., 5\%).
}

To assess the performance of the algorithms, we use \textit{HV} \revision{and \textit{IGD} indicators by following the guideline in~\cite{ali2020quality}. For \textit{HV},} a higher value means a better performance of an algorithm (Section~\ref{sec:experimentDesign})\revision{, while a smaller \textit{IGD} value indicates a better performance}.
With the employed statistical tests,  
based on \revision{the results of indicators} 
achieved by algorithms \textit{A} and \textit{B} on the same prioritization problem, algorithm \textit{A} outperforms algorithm \textit{B} only if \Atwelve is greater than 0.5 \revision{for \textit{HV}, and \Atwelve is less than 0.5 for \textit{IGD}}. 
The difference is considered significant if the \textit{p}-value is less than 0.05. 
Moreover, to demonstrate the performance of the algorithm compared to other algorithms on the different subject systems,
for each problem with each time budget, 
we calculate a \textit{Rank} value for each algorithm 
with Algorithm~\ref{alg:RankAlgorithm}.
Note that a higher \textit{Rank} value means better performance.
\begin{figure}[!t]
\input{tables/algRank}
\end{figure}
With rank values, we further define \textit{Confidence}, which computes the percentage of a \MOSA being better than the others as:
\begin{equation}\label{eq:confidence}
    Confidence_{j} = \frac{Rank_{j}}{\sum^{n}_{i=1}{Rank_{i}}} \times100\%,\;0<j\le n
\end{equation}
To compare the prioritization problems for RQ4, we defined \anou (Section~\ref{sec:experimentDesign}). 
The comparison analysis on the problems was the same as performed on the algorithms, but the metric differed (i.e., \anou for problems and \textit{HV} for algorithms).


To answer RQ3 and RQ4, we employed Spearman's rank correlation coefficient~\cite{spearman1987proof} to examine whether there is a correlation between two variables (e.g., time budgets and \anou) for each problem in RQ3). 
We report the correlation coefficient ($\rho$) and significance of correlation ($p$-value) with a significance level of 5\%.
$p$-value determines whether two variables are monotonically related, i.e., a monotonic relationship between the time budgets and \anou can be identified only if $p$-value is less than 0.05.
Coefficient ($\rho$) represents the direction and strength of the correlation with its value between -1 and +1.
When a monotonic relationship exists, $\rho > 0$ means a positive correlation between time budget and \anou, i.e., \anou tends to increase as the time budget increases, while $\rho < 0$ suggests that \anou tends to decrease as time budget increases.
Based on the guideline from ~\cite{schober2018correlation
}, $|\rho| > 0.9$ could be considered as a \textit{very strong} correlation between two variables, and $|\rho| <0.1$ indicates the correlation is \textit{negligible}.
%

%
%
%

\subsection{Experiment Results}\label{sec:experimentResult}
In this section, we analyze the experimental results to answer all RQs (Sections~\ref{sec:answersRQ1} -~\ref{sec:answersRQ4}).
Note that detailed statistical results can be found in our online repository\textsuperscript{\ref{foot:link}}.


\subsubsection{Answers for RQ1}
\label{sec:answersRQ1}

To answer RQ1, we conducted comparisons of the four selected algorithms with \textit{RS} for solving the 10 search problems (see Table~\ref{table:tabFitnessFunction}) on the 5 use cases from the 2 subject systems (i.e., the 4 selected use cases for AW and 1 use case for GS), using the 10 time budgets regarding \textit{HV} \revision{and \textit{IGD}}. 
Thus, we have 2000 (i.e., 4 $\times$ 10 $\times$ 5 $\times$ 10) combinations of comparisons.
The Mann-Whitney U test, \revision{the Holm–Bonferroni method}, and the Vargha and Delaney statistics were employed to compare each \MOSA with RS\revision{, and the results of \textit{HV} and \textit{IGD} for each comparison between each \MOSA and \textit{RS} under each time constraint can be found in the online repository\textsuperscript{\ref{foot:link}}}.
%

To summarize the results, in 97.45\% (1949/2000) of the cases regarding \textit{HV} \revision{and 81.4\% (1628/2000) of the cases regarding \textit{IGD}}, the selected \MOSA{s} significantly outperformed \textit{RS} since all \Atwelve values were greater than 0.5 and \textit{p-value}s were less than 0.05. In 0.45\% (9/2000) of the cases \revision{in terms of \textit{HV} and 18.6\% (372/2000) in terms of \textit{IGD}}, we observed no significant differences between the \textit{RS} and the selected \MOSA{s} since \textit{p-value}s were greater than 0.05. 
\revision{Regarding the cases that \textit{RS} outperformed \MOSA{s}, in terms of \textit{HV}, we found 2.1\% (42/2000) of the cases, which are only observed in \textit{AW4} (18 cases out of 400) and \textit{GS1} (24 cases out of 400) when using \textit{SPEA2}. Regarding \textit{IGD}, there are no downside cases, i.e., 0\% (0/2000).}
%
\result{
 RQ1: The prioritization problems are complex, hence warranting the use of \MOSA{s} to solve them since for 97.9\% \revision{with \textit{HV} and 100\% with \textit{IGD}} of the time \MOSA{s} either significantly outperformed \textit{RS} (97.45\% \revision{with \textit{HV} and 81.4\% with \textit{IGD}}) or there were no significant differences observed (0.45\% \revision{with \textit{HV} and 18.6\% with \textit{IGD}}). 
}

\subsubsection{Answers for RQ2}\label{sec:answersRQ2}

With RQ2, we aim to find the best MOSA among \MOSA{s} for each prioritization problem with each time budge, with the Kruskal-Wallis Test, Mann-Whitney U Test, Vargha and Delaney statistics, \revision{and Holm–Bonferroni method}. \revision{Detailed comparison results of the 10 prioritization problems on the five use cases in the 10 time budgets with \textit{HV} and \textit{IGD}} can be found in our online repository\textsuperscript{\ref{foot:link}}.

Table~\ref{table:tabBestMOSA_HV_IGD_comb} shows the overall best \MOSA for each problem for each time budget based on the best algorithms of all the use cases. 
\revision{The table is summarized by following two steps. First, for each use case, we select algorithms that performed the best regarding \textit{HV} and/or \textit{IGD} for each problem under each time budget. If one of the selected algorithms achieved the best regarding both \textit{HV} and \textit{IGD}, we opt for it as the best. For instance, for \probFirst within TB10 on \textit{AW1}, both \textit{NSGA-II} and \textit{SPEA2} achieve the best \textit{HV}, and only \textit{SPEA2} achieves the best \textit{IGD}, then \textit{SPEA2} is recognized as the best for solving \probFirst within TB10 on \textit{AW1}. 
If there exist distinct best algorithms in terms of \textit{HV} and \textit{IGD}, we then follow the guideline in~\cite{ali2020quality} to employ the recommended indicator for deciding the best. For instance, for \probSixth within TB40 on \textit{AW2}, \textit{MoCell} achieves the best \textit{HV} and \textit{SPEA2} achieves the best \textit{IGD}. As \textit{IGD} is the recommended indicator in the pair comparison of \textit{MoCell} and \textit{SPEA2} according to the guideline, we then identify \textit{SPEA2} as the best algorithm for solving \probSixth within TB40 on \textit{AW2}. 

Second, based on the summary of the best algorithm for each problem within each time budget on each use case, we select the algorithm that achieves the highest number of best results among the use cases as the best one. For instance, \textit{SPEA2} is selected as the best algorithm for solving \probFirst within TB10 as \textit{SPEA2} is recognized as the best on \textit{AW2}, \textit{AW3}, \textit{AW4} and \textit{GS1} in solving \probFirst within TB10. 
}

\begin{table}[!t]
\caption{\revision{Summary of the best \MOSA{(s)} for each Problem with each time budget \textit{- RQ2}. 
Green--MOCell; Blue--SPEA2; Yellow--Multiple \MOSA{s}.}}
\label{table:tabBestMOSA_HV_IGD_comb}
\centering
\resizebox{0.492\textwidth}{!}{
\input{tables/tabBestMOSA_HV_IGD_comb}
}
\end{table}

\revision{Based on Table~\ref{table:tabBestMOSA_HV_IGD_comb}, \textit{SPEA2} achieves the best results in 96 out of 100 configurations, i.e., 10 problems under 10 time constraints.
However, users might have their own preferences and available time budget to execute tests.
We provide this recommendation table (Table~\ref{table:tabBestMOSA_HV_IGD_comb}) for the users}
to select an algorithm for a particular test prioritization problem for a particular time budget. 
For example, if one has a particular interest in solving \probNineth, i.e., minimizing \textit{PET} and maximizing \textit{PTR}, \textit{PUS}, and \textit{PUU} with any time budget (10\% to 100\%), we recommend using \textit{SPEA2} since it consistently performed the best for all the use cases and for all the time budgets.

\result{
RQ2: \revision{We recommend \textit{SPEA2} as it achieved the best performance in 96 out of 100 configurations .} 
However, \revision{as users could have their own interests and available budget to perform test execution}, a recommendation table is derived to guide the algorithm selection for a specific problem within an available time budget.
}

\subsubsection{Answers for RQ3}\label{sec:answersRQ3}

RQ3 studies the impacts of 
various time budgets for each problem with the best algorithm obtained in RQ2. We conducted a correlation analysis on \anou achieved with various time budgets (i.e., TB10 -- TB100) using the Spearman Correlation Coefficient.
Results (i.e., $\rho$ and $p$-value) are reported in Table~\ref{table:tabSpearmAllCSs}.

\begin{table}[!t]
\caption{\revision{Results of the Spearman Correlation Coefficient between various time budgets and \anou with the best \MOSA{(s)} for each problem on all case studies \textit{- RQ3}}}
\label{table:tabSpearmAllCSs}
\centering
\resizebox{.49\textwidth}{!}{
\input{tables/tabSpearm_casestudies_HV_IGD}
}
\resizebox{.48\textwidth}{!}{
\begin{threeparttable}
\begin{tablenotes}
\footnotesize \item[*] Note that an underlined \underline{value} means that the correlation is not statistically significant (i.e., $p$-value \textgreater{} 0.05; otherwise, significant (i.e., $p$-value $<$ 0.05). 
\end{tablenotes}
\end{threeparttable}}
\end{table}
Based on the results in Table~\ref{table:tabSpearmAllCSs}, 
for all of the 10 problems on the four use cases, except for \probEighth on GS1, a monotonic relationship exists between time budgets and \anou (i.e., $p$-value $<$ 0.05). 
Regarding the overall tendency,
for AW1 and AW2, the monotonic relationship is positive (i.e., $\rho$ $>$ 0), meaning that as the time budget increases, \anou observed by executing tests as the sequence prioritized by \approach tends to increase.
For AW4 and GS1, the monotonic relationship between is positive in most problems, except for \probThird on GS1, \probSixth \revision{on AW4 and GS1}, and \probTenth on GS1.
However, the correlation coefficient of negative cases is \textit{negligible}, i.e., $|\rho| <$ 0.1.
Results obtained in AW3 are different from the other four use cases, for all of the problems, the monotonic relationship between time budgets and \anou is negative, i.e., $\rho$ $<$ 0. It indicates that as the time budget increases, the \anou tends to decrease with the prioritized sequence to execute the tests.

To identify the TB strategy, Table~\ref{table:tabBestTBtoANOU_HV_IGD} demonstrates the best TB, which achieves the best \anou for each problem on each use case. 
The results showed that the TB100 strategy performed best with all problems on AW1 and AW2.
For AW4 and GS1, except for four problems that contain \anu, the TB100 strategy achieved the best in the other six problems.
For AW3, the low time budget (i.e., TB10, TB20\revision{, and TB30}) obtained the best results among all of the problems.
The results of the best TB strategy are consistent with the results of the overall tendency for each use case (Table~\ref{table:tabSpearmAllCSs}), i.e., the positive coefficient results in the greater TB as the best while the negative coefficient results in the lower TB as the best.

\begin{table}[!t]
\caption{\revision{
Summary of the best TB strategy for each problem on each use case \textit{- RQ3}}
}
\label{table:tabBestTBtoANOU_HV_IGD}
\centering
\resizebox{0.49\textwidth}{!}{
\input{tables/tabBestTBtoANOU_HV_IGD}
}
\end{table}

\begin{figure}
    \centering
    \includegraphics[width=0.7\linewidth]{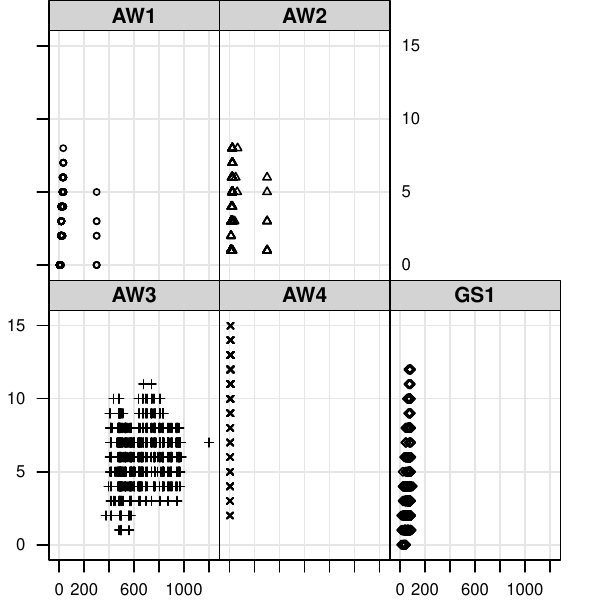}
    \caption{A scatter plot of execution time ($x$ axis in seconds) of each test case versus a number of uncertainties observed per test case (i.e., \nou in $y$ axis) for each use case \textit{- RQ3}}
    \label{fig:tcet}
\end{figure}

Based on the above results, the performance of TB strategies varies from use case to use case and from problem to problem.
To further investigate performance with various TB, for each use case, we reported a scatter plot of execution time spent by each test case along with its observed uncertainties (i.e., \nou) as shown in Figure~\ref{fig:tcet}, and detailed statistics (i.e., average, maximum, minimum and standard deviation) of execution time as shown in Table~\ref{table:et}.
In this table, we also report the average time (seconds) to observe an actual occurrence of uncertainty, i.e., Avg(ET/NOU). 
Based on the results, compared to other use cases, we found that test cases in AW3 spent more time, and differences among test cases are greater.
The time cost to observe an uncertainty in AW3 (i.e., 117.40 seconds on average) is also much more than in other use cases.
Thus, with an increase in time budget, the efficiency metric \anou would probably decrease.
This may explain the negative correlation result between time budget and \anou in AW3, while a positive correlation exists in other use cases in most cases.

\begin{table}
\small
\centering
\caption{Descriptive statistics of the execution times (in seconds) of test cases to prioritize for each use case, and the average time (in seconds) needed to observe an uncertainty (i.e., Avg(ET/NOU)) \textit{- RQ3}}
\label{table:et}
\resizebox{.99\linewidth}{!}{
\input{tables/et}
}
\end{table}

\begin{figure}
    \centering
    \includegraphics[width=0.7\linewidth]{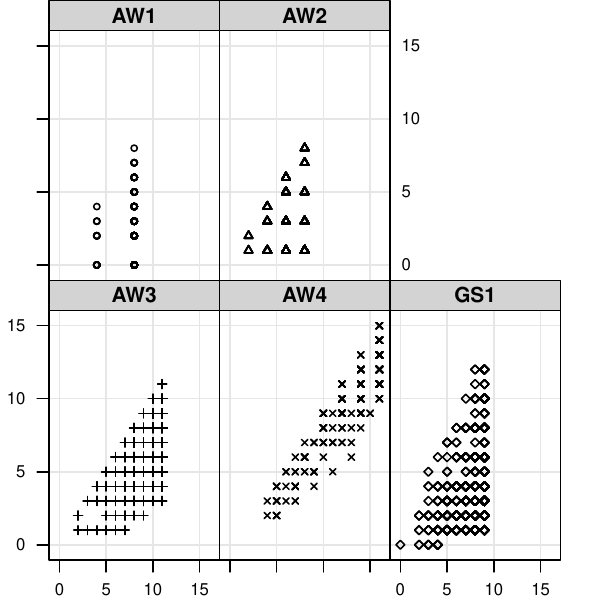}
    \caption{A scatter plot of a number of subjective uncertainties (i.e., $x$ axis is \nsu) of each test case versus a number of uncertainties observed per the test case (i.e., $y$ axis is \nou) for each use case \textit{- RQ3}}
    \label{fig:tcnu}
\end{figure}

Based on results of correlation analysis (Table~\ref{table:tabSpearmAllCSs}) and best TB (Table~\ref{table:tabBestTBtoANOU_HV_IGD}), in AW4 and GS1, 
the problems which contain \anu achieved different best TB strategies than other problems. 
The \anu objective is derived from a number of subjective uncertainties (i.e., \nsu) of test cases based on their prioritized sequence.
To analyze \nsu with \anou in various use cases,
we reported a scatter plot of \nsu specified in each test case versus its observed uncertainties (i.e., \nou) as shown in Figure~\ref{fig:tcnu}, and performed a Spearman Correlation analysis on \nsu and \nou of test cases for each use case (Table~\ref{table:spearman_nu_nou}).
With the results, we found that in AW3, AW4 and GS1, a greater \nsu trends to observe more \nou, while such a correlation is not significant in AW1 and AW2.
Thus, in AW3, AW4 and GS1, the problems with \anu objective would result in a higher chance to prioritize tests that are capable of observing more uncertainties in the front positions of the execution sequence.
For AW4 and GS1, since there does not exist much difference in execution time (see $\sigma$ in Table~\ref{table:et}), test cases in the front positions of the sequence would probably achieve higher \anou than ones in the rear positions.
Thus, this might explain that the less time budget (i.e., less than TB100) achieves the best \anou for the problems that contain \anu in AW4 and GS1.
In Table~\ref{table:spearman_nu_nou}, we also report the average time to target a specified subjective uncertainty, i.e., Avg (ET/NU).
For all of the use cases, the actual time spent to observe an uncertainty is longer than expected, i.e., Avg(ET/NOU) $>$ Avg (ET/NU).
Thus, if test cases of a use case take time to target a subjective uncertainty, such as more than 1 minute in AW3, it would probably take more time to observe an uncertainty.
Therefore, for use cases that take more time to target a subjective uncertainty \revision{(i.e., Avg(ET/NU) $>$ 60s)} as the efficiency would probably decrease over time, we would recommend less time budget as the TB strategy, i.e., TB10/TB20\revision{/TB30}.

\textbf{\begin{table}
\small
\centering
\caption{Results of the Spearman Correlation Coefficient between \nsu and \nou of test cases for each use case \textit{- RQ3}}
\label{table:spearman_nu_nou}
\input{tables/spearman_nu_nou}
\begin{threeparttable}
\begin{tablenotes}
\footnotesize \item[*] Note that an underlined \underline{value} means that the correlation is not statistically significant (i.e., $p$-value \textgreater{} 0.05); otherwise significant. 
\end{tablenotes}
\end{threeparttable}
\end{table}}

Based on the above analyses, we found that TB100 achieved the best in most cases. 
For cases that spend more execution time and have fewer subjective uncertainties, we would recommend fewer TB strategies (such as AW3).
If test cases in a use case spend less time, the best TB strategy would highly depend on use cases and problems such as AW1 $vs.$ AW4, we cannot conclude a clear recommendation currently. However, if the time budget is not a practical constraint, we recommend TB100.
\result{
RQ3: Significant correlations between time budget and \anou are observed in all five use cases, but the direction and strength are use case by use case.
TB100 (i.e., executing all tests) is recommended when the time budget is not a practical constraint; otherwise, we recommend strategies requiring smaller TB (i.e., TB10/TB20\revision{/TB30}), especially for cases with few uncertainties specified.
}

\subsubsection{Answers for RQ4}\label{sec:answersRQ4}

RQ4 studies efficiency among 10 \uncertaintyaware test case prioritization problems.
The efficiency of each problem is measured with \anou achieved by the best \MOSA using TB100 (selected in RQ3). Detailed configurations for each problem on each use case \revision{can be found in the online repository\textsuperscript{\ref{foot:link}}.}

\begin{figure*}
     \centering
     \begin{subfigure}{0.19\textwidth}
         \centering
         \includegraphics[width=\textwidth]{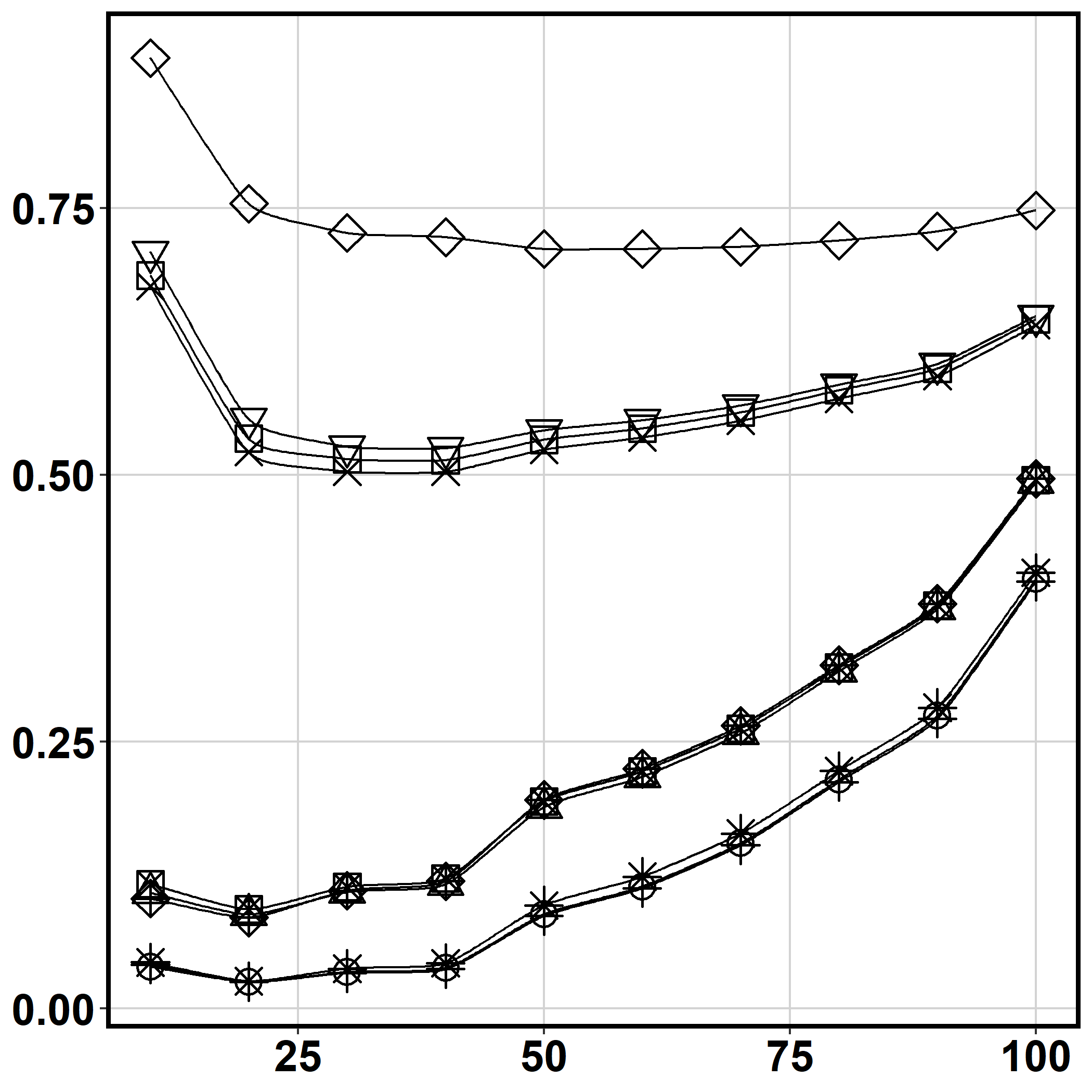}
         \caption{AW1}
         \label{fig:problemsAW1_comb}
     \end{subfigure}
     \hfill
     \begin{subfigure}{0.19\textwidth}
         \centering
         \includegraphics[width=\textwidth]{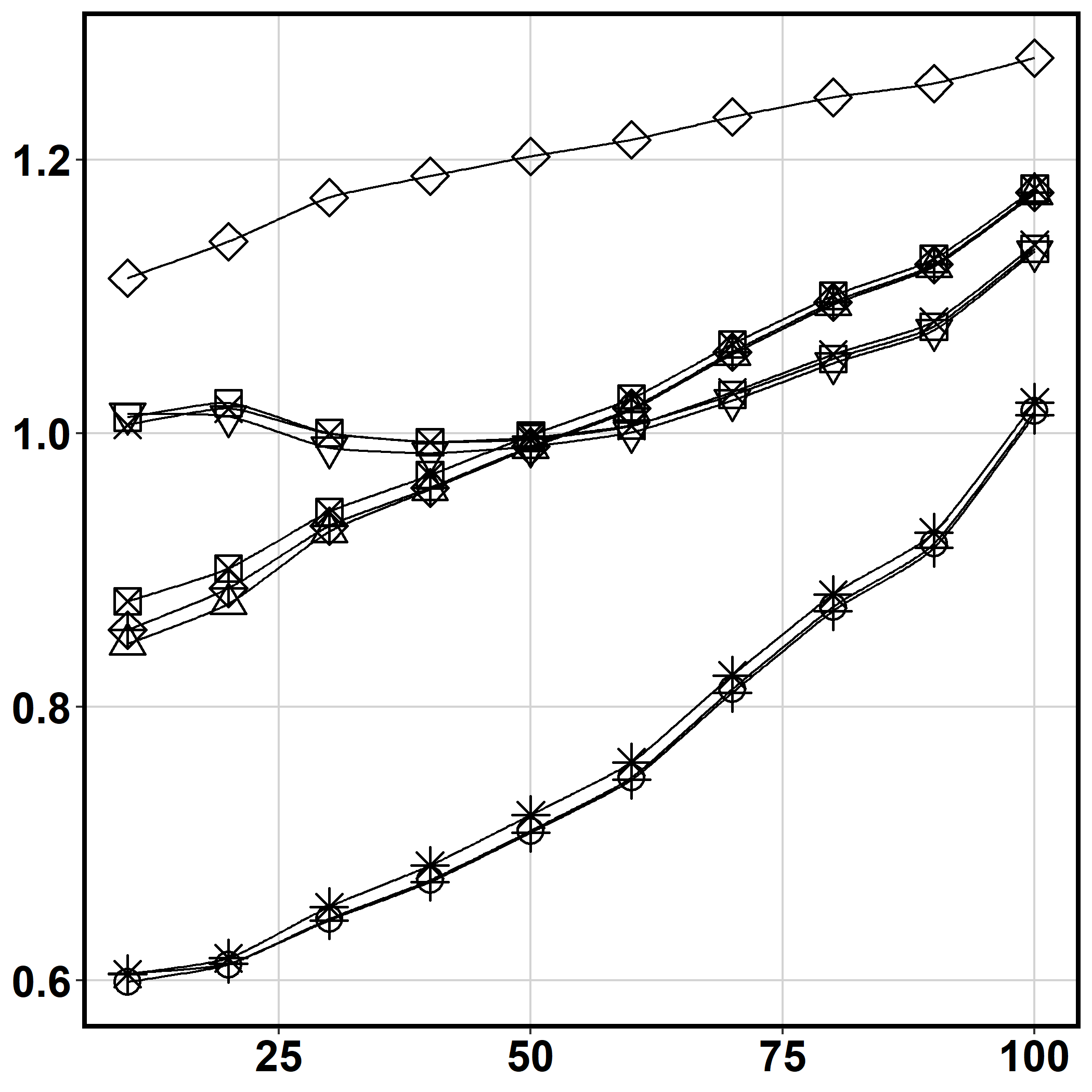}
         \caption{AW2}
         \label{fig:problemsAW2_comb}
     \end{subfigure}
     \hfill
     \begin{subfigure}{0.19\textwidth}
         \centering
         \includegraphics[width=\textwidth]{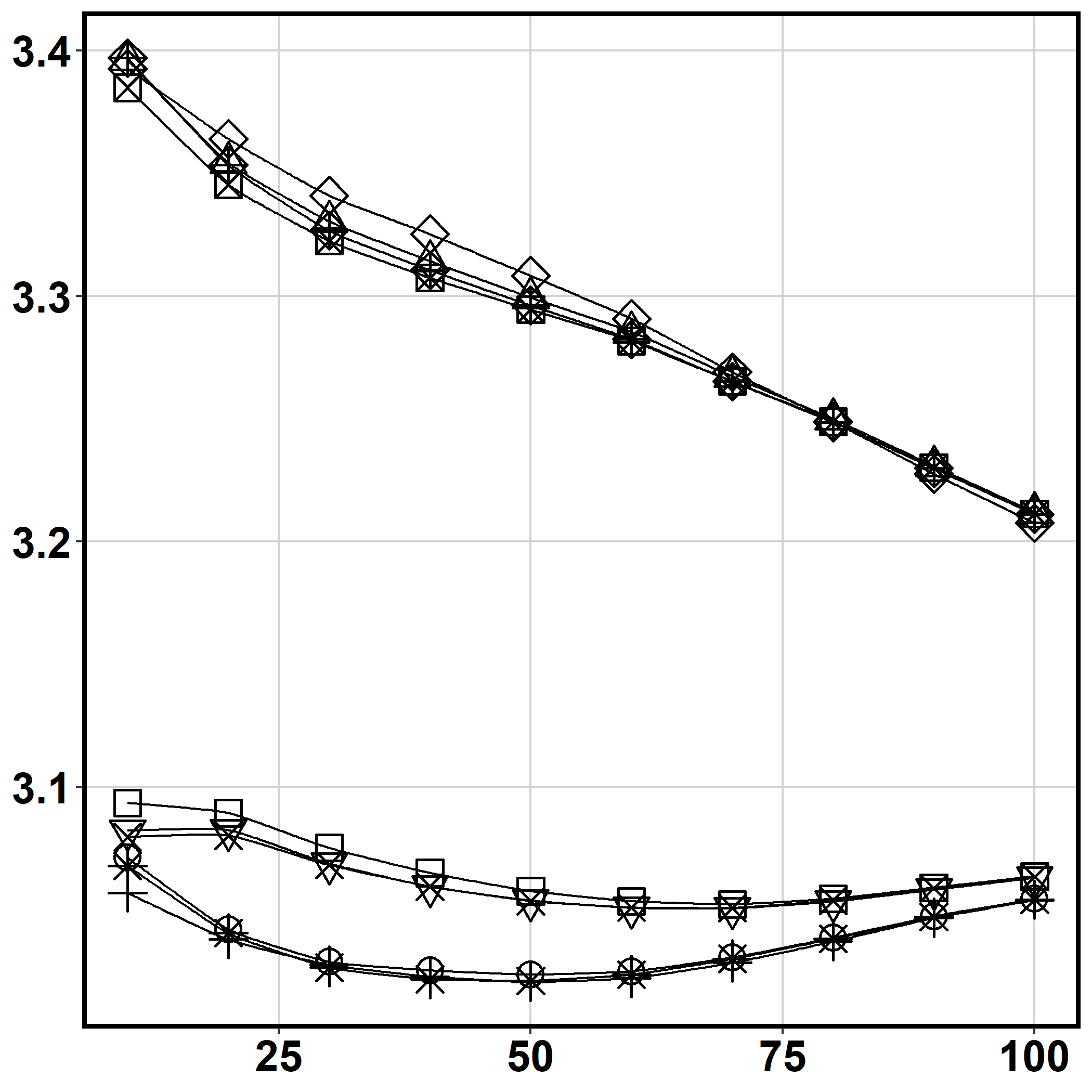}
         \caption{AW3}
         \label{fig:problemsAW3_comb}
     \end{subfigure}
     \hfill 
    \begin{subfigure}{0.19\textwidth}
         \centering
         \includegraphics[width=\textwidth]{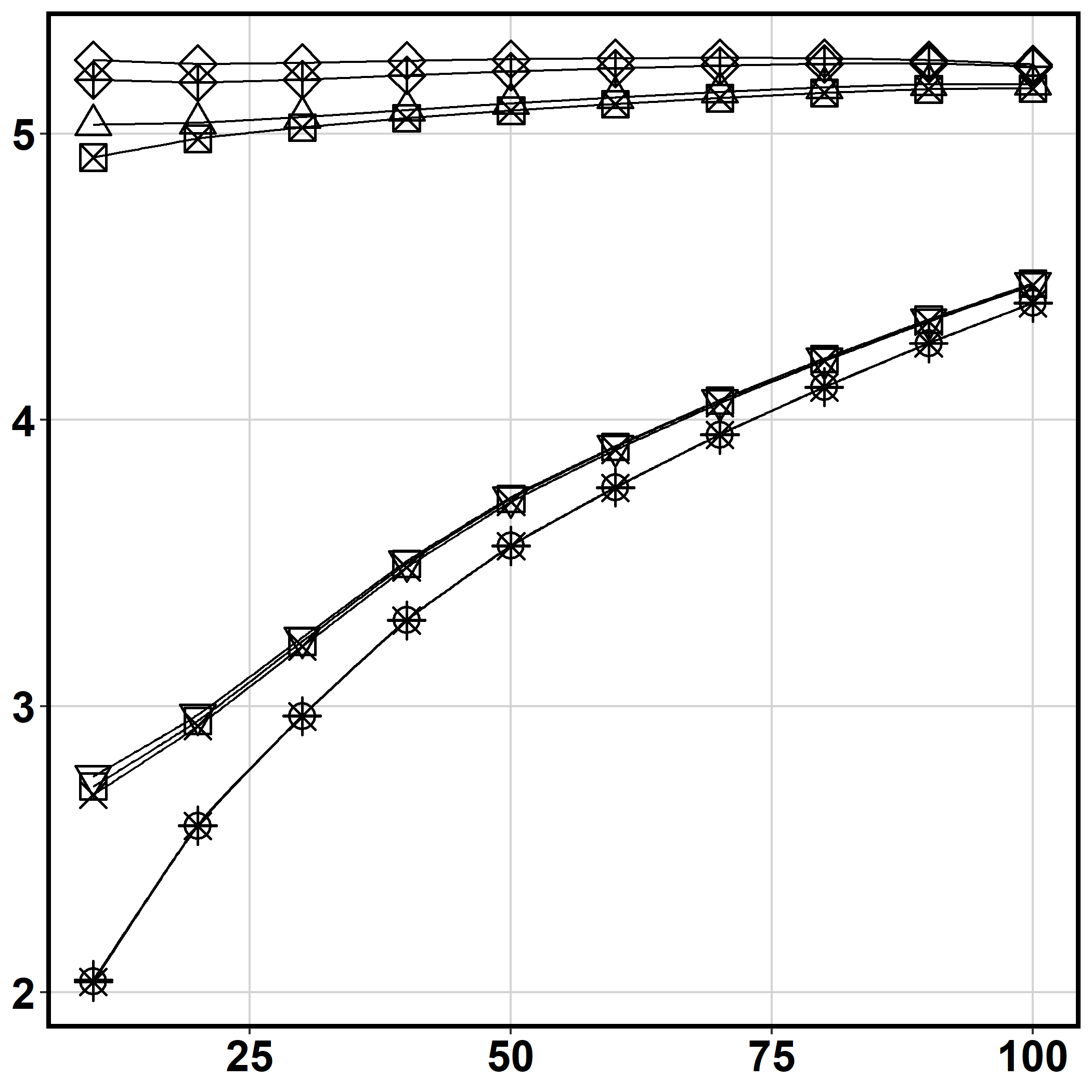}
         \caption{AW4}
         \label{fig:problemsAW4_comb}
     \end{subfigure}
     \hfill
     \begin{subfigure}{0.19\textwidth}
         \centering
         \includegraphics[width=\textwidth]{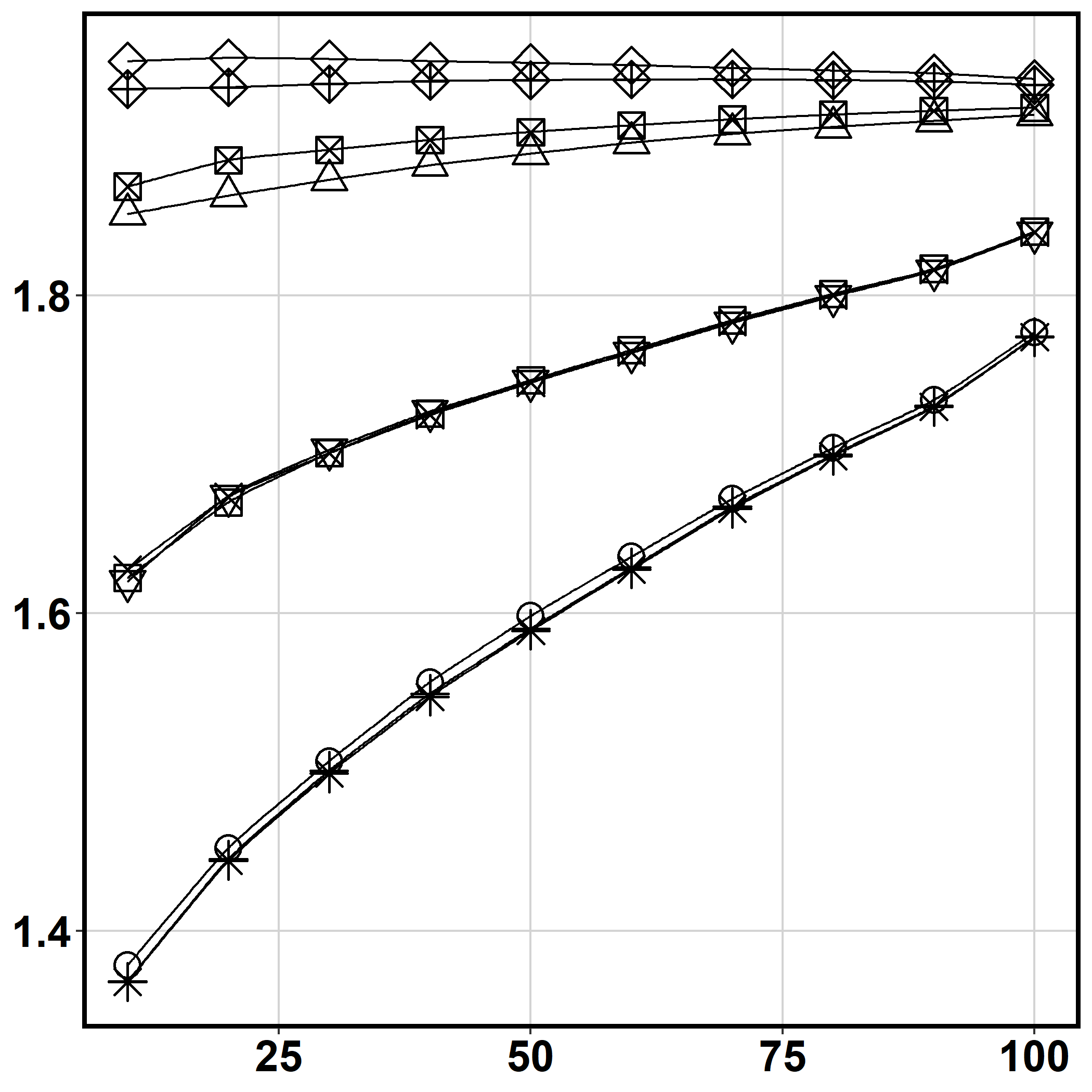}
         \caption{GS1}
         \label{fig:problemsGS1_comb}
     \end{subfigure}
     \hfill 
     \begin{subfigure}{\textwidth}
         \centering
         \includegraphics[width=\textwidth]{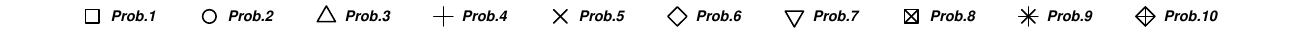}
         \label{fig:problemslegend_comb}
     \end{subfigure}
     \vspace{-0.95cm}
     \caption{\revision{Average \anou ($y$ axis) achieved with the best \MOSA{s} reported at every 10\% used time budget ($x$ axis) with the TB100 strategy for all of the 10 prioritization problems on each use case \textit{- RQ4}}}
        \label{fig:problemsAll_comb}
\end{figure*}

To compare the problems, we conducted comparative analyses and reported results of \textit{Rank} (see Algorithm~\ref{alg:RankAlgorithm}) and \textit{Confidence} (see Equation~\ref{eq:confidence}) in Tables~\ref{table:tabProblemRank_comb} and~\ref{table:tabProblemConfidence_comb} respectively.
More detailed results of the Kruskal-Wallis Test, Mann-Whitney U Test, Vargha and Delaney effect size, \revision{and Holm–Bonferroni method} can be found in our online repository\textsuperscript{\ref{foot:link}}.
Regarding the results of \textit{Rank} and \textit{Confidence}, we found that \probSixth performed the best (i.e., the highest rank value) in four out of the five use cases (i.e., except AW3) and performed the third best (inferior to \textit{Prob} 3, 8 and 10) in AW3.
In addition, \probSixth achieved the overall best performance among the five use cases with \textit{Rank} and \textit{Confidence} metrics (see \textit{All} in Tables~\ref{table:tabProblemRank_comb} and~\ref{table:tabProblemConfidence_comb}).

\begin{table}[!t]
\caption{\revision{Rank results of each problem on each use case with \anou \textit{- RQ4}. Note that a greater value means a better performance, and the grey cells are the best.}}
\label{table:tabProblemRank_comb}
\centering
\resizebox{0.45\textwidth}{!}{
\input{tables/tabProblemRank_comb}
}
\end{table}
\begin{table}[!t]
\caption{\revision{Confidence results of each problem on each use case with \anou \textit{- RQ4}. Grey cells highlight the best.}}
\label{table:tabProblemConfidence_comb}
\centering
\resizebox{0.45\textwidth}{!}{
\input{tables/tabProblemConfidence_comb}
}
\end{table}

In order to provide more details on the performance of the 10 problems, Figure~\ref{fig:problemsAll_comb} plots the average
\anou over time spent by executing tests with the sequence prioritized by each problem, and the results are reported at every 10\% time budget under the TB100 strategy. 
In terms of average \anou, prioritized solutions achieved by \probSixth outperformed solutions by other problems throughout the entire execution in four out of the five use cases except for AW3.
\revision{Regarding AW3, a group of problems, i.e., \textit{Prob.}3, 6, 8, and 10 achieved the best, and the differences among them are modest.
Thus, \probSixth can be recognized as the best among all five use cases.}

To investigate subjective uncertainty objectives defined in the problems responding to \anou, for each use case, we conducted Spearman's rank correlation coefficient for each subjective measure for a test (i.e., \um, \nsu, \nsuu, and \usp) with a number of observed uncertainties by the test (i.e., \nou).
In addition, we analyzed the correlation coefficient between each pair of subjective measures to study uncertainty properties in the tests being prioritized.
Results are shown in Table~\ref{table:spearman_um_nu_nou}.
Regarding \nsu, \nsuu, and \usp responding to \nou (i.e., \nou$\sim$\nsu, \nou$\sim$\nsuu, and \nou$\sim$\usp), there does not exist a significant correlation between the three measures and \nou in AW1 and AW2, and a positive correlation is observed in AW3, AW4 and GS1. 
In terms of the significant correlations,
comparing with \nsuu and \usp, \nsu obtained a higher positive strength (see $\rho$) signifying that a greater \nsu has a higher chance to lead to a higher \anou. 
In addition, with the correlation results of \nsu, \nsuu, and \usp,
we found that they are all positively correlated (see \nsu$\sim$\nsuu, \nsu$\sim$\usp, and \nsuu$\sim$\usp).
Thus, there might be a lack of interactions between any pair of the three measures.
Moreover, since \nsu (referring to \anu) might be superior to \nsuu (referring to \textit{PUU}) and \usp (referring to \textit{PUS}), this might explain that the problems containing \anu are consistently better than the problems containing\textit{PUU} or \textit{PUS} in all use cases.
Another piece of evidence is that the four problems containing \anu are consistently the best in four out of five use cases (except for AW1). 

In terms of \um, there does not exist a clear correlation between \um and \nou as (1) in AW1--4, $p$-value is greater than 0.05, and (2) in GS1, $p$-value is less than 0.05 but $\rho$ is low (i.e., 0.11).
Regarding the correlation of \um with \nsu, \nsuu, and \usp, results are consistent for the three measures, i.e., $p$-value is less than 0.05 and $\rho$ is negative in all of the use cases.
It indicates that, for tests being prioritized, a test with more subjective uncertainties (i.e., a greater \nsu) or more unique subjective uncertainties (i.e.,  a greater \nsuu) or covering more uncertainty spaces (i.e.,  a greater \usp) tends to have a lower uncertainty measure (i.e., \um).
\um for a test is calculated based on a minimum value of \um of all uncertainties it contains, then a test with more (unique) uncertainties or covering more uncertainty space would result in a higher chance to include an uncertainty with a lower \um.
This might be a reason for the negative correlation of \um with \nsu, \nsuu and \usp.
As the correlation results, \um and \nou are not significantly correlated, but
with the results of problems that contain \aum, we found that the best four in AW1 are the problems that contain \aum.
This might be due to a non-significant correlation between \nsu and \nou in AW1, and comparing \nsu and \um in this use case, \um might better guide to observing the uncertainties.
However, combining \nsu with \um, they could interact with each other, e.g., if tests have the same \nsu, \um might be possible to distinguish the tests that have a greater \um in the front positions.
Such prioritized tests might result in better performance, as \probSixth achieved the overall best.
\result{
RQ4: An \uncertaintyaware prioritization problem with maximizing \aum and maximizing \anu (i.e., \probSixth) achieved the overall best performance in terms of efficiency in observing uncertainties. 
}

\begin{table}
\small
\centering
\caption{Results of the Spearman Correlation Coefficient for each pairs of \nou, \um, \nsu, \nsuu, and \usp
of test cases for each use case \textit{- RQ4}}
\label{table:spearman_um_nu_nou}
\resizebox{.99\linewidth}{!}{
\input{tables/spearman_um_nu_nou}
}
\end{table}

\begin{figure}
    \centering
    \includegraphics[width=0.99\linewidth]{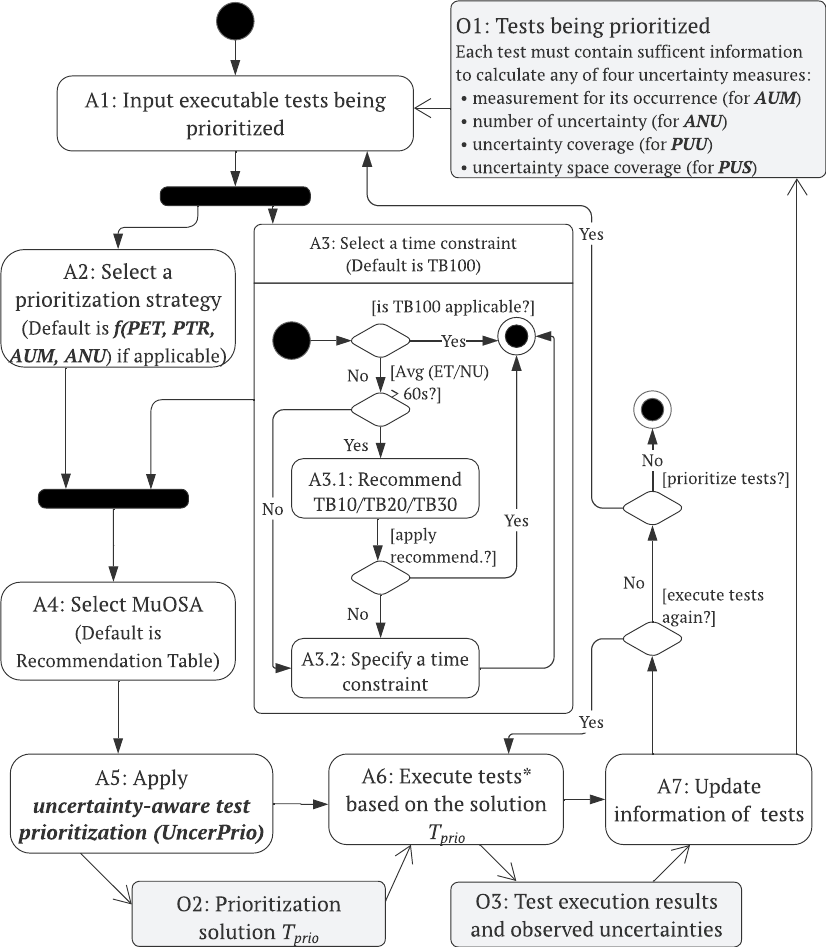}
    \raggedright \footnotesize \revision{* Test execution environment should be controllable/accessible in order to enable/evaluate indeterminacy sources.}
    \caption{\revision{Step-wise guideline of applying \approach}}
    \label{fig:guideline}
\end{figure}

\revision{
\subsection{Applying \approach}
\approach is designed to optimize the test execution process specific to SUTs facing uncertainty (e.g., unpredictable operating environment). As shown in Figure~\ref{fig:guideline}, first, applying \approach requires a set of \textit{executable tests} with sufficient information for calculating the four uncertainty measures (\textbf{A1}). 
Note that \approach was implemented as a series of uncertainty-wise solutions including an uncertainty modeling approach (\textit{UncerTum}) and an automated test generation approach (\textit{UncerTest}), as shown in Figure~\ref{fig:overview}. However, \approach can be independent of them as long as tests have sufficient information as \textit{O1} (e.g., a demo example with JSON provided in our repository\footnote{https://github.com/man-zhang/uncertainty-prioritization/blob/main/example/foo.json}).

Users of \approach can select an \textit{applicable} strategy, in the sense that required information is available for calculating required uncertainty measures, to prioritize tests (\textbf{A2}) based on preferences, though \approach's default strategy is \probSixth, as it achieves the overall best performance in observing uncertainties in our experiment.  
Considering the available time budget and repetition times of test executions, a user can decide on a time constraint (\textbf{A3}). \approach's default time budget is TB100, which is set based on our experimental results. 
If there is a lack of knowledge about the SUT and its operating environment, we recommend executing tests multiple times to gain (objective) knowledge based on execution results. However, we do not have a concrete recommendation on the number of repetitions, as an empirical study involving industry practitioners is needed in the future to derive such information.
With the specified strategy and time constraint, we recommend an \MOSA by referring to Table~\ref{table:tabBestMOSA_HV_IGD_comb} (\textbf{A4}). However, a user can specify one to use if needed. 

We implemented \approach as an open-source prototype tool, which is online available\textsuperscript{\ref{foot:link}}. The tool can be used (\textbf{A5}) via a command line as its user interface and is accompanied by documentation and an example demonstrating how to use it. Guided by solutions produced by the tool (\textit{O2}), users can conduct test execution on their test execution infrastructures (\textbf{A6}) and obtain test execution results (\textit{O3}). 
To ease the introduction of indeterminacy sources and conveniently observe their occurrences, we recommend making the test infrastructure controllable (e.g., manipulating the test environment) and accessible (e.g., retrieving statuses of the test environment). For instance, in our experiment, for testing \textit{GS1}, an available test infrastructure facilitates accessing statuses of the SUT in real time, such as where a device is located. We set up signal shielding at a certain position to simulate a challenging situation whereby the SUT might not be able to receive signals as usual, i.e., introducing an indeterminacy source if the SUT moves to the position where the shielding is located.  

Regarding execution results (\textit{O3}), since we used \textit{UncerTest} to generate tests, which already contain necessary assertions to evaluate the occurrence of uncertainties and indeterminacy sources (see Section~\ref{sec:testResults}).
To be independent of \textit{UncerTest}, users can define their own specifications to observe such occurrences and generate test results after test execution accordingly.
Based on the objective information contained in the test results, information of tests can be updated (\textbf{A7}). For instance, newly observed uncertainties that are \textit{previously unknown} before test execution can be added, and measurements of uncertainty occurrences (e.g., calculating the frequency) can be updated. This requires integrating \textit{subjective} measurements (e.g., \textit{belief degree} with \textit{Uncertainty Theory} in BMs) and \textit{objective} measurements (e.g., \textit{frequency} with \textit{Probability Theory} from the test results). In the future, we plan to investigate different theories and methods for integrating subjective and objective measurements, such as computing a weighted average of the frequency-based probability and belief degree and using Bayesian approaches to update belief degrees with frequency data. 

Once the first round of execution ends, based on the remaining time budget, users can further decide whether to: 1) execute the tests with the current prioritization solution again, 2) generate a new solution with the updated information of tests, or 3) terminate the test execution process.}

\subsection{Threats to Validity}
\label{sec:threatValidity}
\textbf{Internal validity} is related to the parameter settings of the selected \MOSA{s}, i.e., \textit{NSGA-II, MOCell, SPEA2, and CellDE}. 
We chose default settings~\cite{arcuri2011practical, nebro2007design} for these algorithms and
parameter tuning may further improve their performance. 
However, we follow the commonly used guidelines to choose these default values~\cite{arcuri2011practical, sheskin2003handbook}.  

The key threat to \textbf{external validity} is the generalization of the results. We used five different use cases from two different CPS subject systems for the 10 test prioritization problems for 10 different time budgets. No doubt, further experiments with more subject systems can be performed to generalize the results further.

The key threat to \textbf{conclusion validity} is about the randomness of solutions produced by \MOSA{s}~\cite{wang2013minimizing}. We dealt with this threat by repeating the experiments 100 times~\cite{wang2013minimizing} according to guidelines from~\cite{arcuri2011practical}. 
To draw conclusions, results were analyzed with statistical methods, such as
Vargha and Delaney statistics to calculate the effect size, Mann-Whitney U test to determine the significance of results~\cite{yue2013facilitating}, Spearman's rank correlation coefficient to explore correlations of variables, \revision{and the Holm–Bonferroni method as the posthoc analysis to control the family-wise error rate due to multiple comparisons
}.

The key threat to \textbf{construct validity} is about the use of measures to compare \MOSA{s}~\cite{wang2013minimizing}. To avoid incomparability of the measures, we used the same stopping criterion (25000 fitness evaluations~\cite{wang2013minimizing}) for all algorithms to avoid any potential bias in results. Furthermore, we use the same quality indicators, i.e., \textit{HV} and \textit{IGD} to compare all \MOSA{s}.

\section{Related Work}\label{sec:relatedWork}
Test case prioritization orders test cases for execution in terms of their `importance', which can be measured with various testing-related objectives, such as coverage criteria (e.g., branch coverage on code) and fault detection. 
In~\cite{zhang2022test}, Zhang \textit{et al.} sorted all the candidate test cases considering not-yet-covered code units in the previous iteration to select partial test cases for prioritization with a greedy strategy in the context of regression testing. 
Lu \textit{et al.}~\cite{lu2019ant} implemented a coverage-based test case prioritization approach with ant colony optimization to maximize the statement coverage on source code. In addition, due to the increasing use of MBT, model-related coverage objectives are also used for test case prioritization (e.g.,~\cite{marijan2017titan, hierons2020many, parejo2016multi}). For instance, Hierons \textit{et al.}~\cite{hierons2020many} applied the grid-based evolution strategy that included several objectives, including an objective related to pairwise coverage of features (i.e., model elements), to produce test suites and prioritize tests in the context of MBT for software product lines. 
In this paper, the main inputs for generating test cases are test models (i.e., UML state machines), which describe expected system behaviors against which the system shall be tested. To maximize the model coverage (i.e., the coverage of a state machine), all the 10 proposed problems are defined with\textit{PTR} (i.e., percentage of transition coverage with position impact) objective--maximizing the percentage of the total number of unique transitions covered by the prioritized test cases. To cover more uncertainties, we also defined \textit{PUU} and \textit{PUS} to ensure that systems are tested with high coverage of known uncertainties. Such coverage of uncertainties is the novelty of our work.

Another important objective is fault detection in prioritization, whose value, e.g., can be derived based on historical test results. In~\cite{wang2015cost, wang2016practical, arrieta2019search}, the fault detection capability was defined as the rate of successful executions (i.e., executions that found faults) based on the test case execution history. Parejo \textit{et al.}~\cite{parejo2016multi} introduced another objective, \textit{number of faults}, calculated based on fault reports of a previous version of the software under test. To adapt this aspect for our aim (i.e., accelerating observations of uncertainties), we defined \textit{ANOU} (i.e., average normalized number of observed uncertainties with the position impact), indicating that a solution with a higher value of \textit{ANOU} can observe more uncertainties as earlier as possible. However, these data are not always available since existing approaches do not typically consider such uncertainty concerns. Therefore, we used \textit{UM} for a test case and further defined another objective, \textit{AUM} (i.e., average uncertainty measure with the position impact), relying on the knowledge and expertise of domain experts from the subjective perspective. A higher value of \textit{UM} reflects a higher \textit{belief degree} of observing the occurrence of uncertainties included in the test case. Next, a higher value of \textit{AUM} might lead to a higher chance of observing uncertainties with prioritized tests.

As the time budget for executing test cases is always limited, execution time is one widely used objective~\cite{arrieta2016test, arrieta2017employing, arrieta2019search}. For instance, Wang \textit{et al.}~\cite{wang2016enhancing} defined a cost measure referring to the total execution time for enhancing test case prioritization in an industrial setting. In~\cite{shin2018test}, Shin \textit{et al.} proposed to schedule the test execution order considering the time budget for CPS acceptance testing. Arrieta~\textit{et al.}~\cite{arrieta2019search} considered execution time, fault detection, requirements coverage, etc. to solve a search-based test case prioritization problem in the context of testing CPS product lines. 
Similarly, we also take the execution time into account for all the 10 \uncertaintyaware problems and define 10 time budgets for the experiment.

Regarding uncertainty, Zhang \textit{et al.}~\cite{zhang2020uncertainty} implemented a search-based uncertainty-wise approach for requirements prioritization involving the uncertainty in cost overrun, i.e., cost overrun probability. The approach utilized Monte Carlo Simulation to estimate the probability of cost overrun with requirements implementation cost modeled by the triangular distribution. 
In~\cite{shin2018test}, uncertainty was considered a factor that impacts execution time and the risk of hardware damage. With such consideration, execution time and risk were expressed with probability functions as parts of objectives for optimizing acceptance test cases. As demonstrated in the paper, a probability function for execution time can be obtained with the Monte Carlo process, and risk is specified by the related engineer as a single value. Our work aims to prioritize test cases for execution by considering uncertainty. Uncertainty-related information is initially constructed in test models by domain experts (e.g., software testing engineers) from a subjective perspective. With such subjective uncertainty, we defined four objectives, i.e., \textit{AUM}, \textit{ANU}, \textit{PUU} and \textit{PUS} for prioritizing test cases from various uncertainty perspectives (e.g., uncertainty measurement, uncertainty coverage). In addition, with our previous work (i.e., UncerTest), we managed to generate executable test cases based on such models and evaluate occurrences of the unknown uncertainty (not specified in the models) and known uncertainties (specified in the models) with test execution. Thus, objective uncertainty information can be collected accumulatively with test execution. We further defined \textit{ANOU} (referring to the number of observed uncertainties) for prioritizing test cases concerning objective uncertainties. 

In our previous work, we developed an \uncertaintyaware test case generation and minimization solution, named \textit{UncerTest}, as we discussed in Section~\ref{sec:background}. Built on \textit{UncerTest}, we formulated the \uncertaintyaware test case optimization problem and developed a solution accordingly, as presented in this paper. Further, we empirically evaluated the impacts of various mutation and crossover operators for solving the \textit{UncerTest} minimization problems with selected \MOSA. 
\revision{
There exist several existing works for uncertainty-aware testing of CPSs~\cite{wang2018oracles, shin2018test, menghi2019generating, shin2021uncertainty, camilli2021uncertainty} since testing CPSs under uncertainties and also discovering uncertain behaviors are critical to ensure their dependability. However, uncertainty-aware test prioritization has not been studied in the literature. \approach is the first solution that addresses the problem. }

\section{Conclusion and Future Work}\label{sec:conclusion}
In this paper, we proposed a multi-objective \uncertaintyaware and time-aware test case prioritization approach, named \approach, for Cyber-Physical Systems (CPSs). \approach considers four dedicated subjective uncertainty-related objectives, transition coverage, and execution time. 
Such multi-objective problems are then solved with the four commonly used \MOSA{s} (i.e., NSGA-II, MOCell, SPEA2, and CellDE) under 10 time constraints. 
We evaluated the selected \MOSA{s} by comparing with Random Search (RS) using five use cases from two industrial CPSs of varying complexity. 
The results showed that all the \MOSA{s} significantly outperformed RS. 
To investigate the best strategy to prioritize tests in observing uncertainties efficiently, we analyzed the four \MOSA{s}, the 10 time budgets, and the 10 \uncertaintyaware prioritization problems, and observed that the overall best performance was achieved by a problem with combined \aum and \anu~\uncertaintyaware objectives (i.e., \probSixth) using 100\% time budget (i.e., TB100), and a recommended \MOSA for the problem with the TB100 strategy is \revision{SPEA2}.

\revision{In this study, we studied the correlations between two uncertainty measures. In the future, we plan to investigate correlations of more uncertainty measures by defining many objective optimization problems via dedicated empirical studies. We also plan to conduct more case studies to generalize our findings further. Last but not least, we plan to study correlations between uncertainties and bugs and use such correlations to guide practitioners in better identifying and locating bugs.}

\ifCLASSOPTIONcompsoc
  \section*{Acknowledgments}
\else
  \section*{Acknowledgment}
\fi

This work is supported by the Co-tester project (No. 314544), funded by the Research Council of Norway. This work was also supported by the European Research Council (ERC) under the European Union’s Horizon 2020 research and innovation program (grant agreement No 864972).

\ifCLASSOPTIONcaptionsoff
  \newpage
\fi



\bibliographystyle{IEEEtran}
\bibliography{Reference}
%



%

\end{document}

%% file: tables/tabTestCaseAttribute.tex
\begin{threeparttable}
\begin{tabular}{|p{0.6in}|p{0.5in}|p{2.5in}|}
    \hline
    \textbf{Phase} & \textbf{Attribute} & \textbf{Definition}\\
    \hline
    
    Test \quad Case & \textit{TR}$(t)$ & Set of unique transitions covered by test case $t$: \textit{TR}$(t) = \{{tr}_{1}, {tr}_{2}, \dots ,{tr}_{ntr_{t}}\}$, where ${ntr}_{t}$ is the total number of the transitions.\\ \cline{2-3}
    Generation & \textit{Us}$(t)$ & Multiset of uncertainties covered by $t$: \textit{Us}$(t)=\{u_1, u_2, \dots ,u_{nu_t}\}$, where $nu_{t}$ is the total number of the uncertainties.\\ \cline{2-3}
     & \textit{UU}$(t)$ & Set of unique uncertainties covered by $t$: \textit{UU}$(t)=\{u_1, u_2, \dots ,u_{nuu_t}\}$, where ${nuu}_{t}$ is the total number of the uncertainties.\\ \cline{2-3}
    & \textit{USP}$(t)$ & Set of uncertainty spaces covered by $t$: \textit{USP}$(t)=\{usp_1, usp_2, \dots ,usp_{nusp_t}\}$, where ${nusp}_{t}$ is the total number of the uncertainty spaces.\\ \cline{2-3}
    & \textit{UM}$(t)$ & Uncertainty measurement of $t$. For instance, \textit{UM}$(t)=0.8$ means that (1) with \textit{Uncertainty Theory}: the modeler has 80\% confidence that the execution of test case $t$ will pass; (2) with \textit{Probability Theory}: the test case $t$ will pass with a probability of 80\% based on its execution history~\cite{zhang2019uncertainty}. Note that \textit{Uncertainty Theory} was applied in \textit{UncerTest}.\\ \hline 
    Test \quad Case \quad Execution & \textit{ET}$(t)$ & One-time execution time of test case $t$. The average execution time for multiple executions of $t$: $AET(t)=\frac{\sum^{n}_{k=1}{ET^{k}(t)}}{n}$, where $n$ is the times of executions and $ET^{k}(t)$ is the time for the $k$th time execution. \\ \hline
\end{tabular}


\end{threeparttable}

%% file: tables/tabFitnessFunction.tex
%
    
    

\begin{tabular}{|l|c| l | c|}
    \hline
    \textbf{Prob.} & \textbf{Fitness Function} & \textbf{Prob.} & \textbf{Fitness Function}\\
    \hline
    
    1 & \textit{f(PET,PTR,AUM)} & 6 & \textit{f(PET,PTR,AUM,ANU)}\\ \hline
    2 & \textit{f(PET,PTR,PUS)} & 7 & \textit{f(PET,PTR,AUM,PUU)}\\ \hline
    3 & \textit{f(PET,PTR,ANU)} & 8 & \textit{f(PET,PTR,PUS,ANU)}\\ \hline
    4 & \textit{f(PET,PTR,PUU)} & 9 & \textit{f(PET,PTR,PUS,PUU)}\\ \hline
    5 & \textit{f(PET,PTR,AUM, PUS)} & 10 & \textit{f(PET,PTR,ANU,PUU)}\\ \hline
\end{tabular}

%
%

%% file: tables/expDesign.tex
\begin{tabular}{|c|c|c|c|ll|c|l|c|}
    \hline
    \textbf{RQs} & \textbf{Task} & \textbf{Problem} & \textbf{\begin{tabular}[c]{@{}c@{}}Time\\ Budget (TB)\end{tabular}} & \multicolumn{2}{c|}{\textbf{Algorithm}} & \textbf{Metric} & \multicolumn{1}{c|}{\textbf{Statistical Analysis}} & \textbf{\begin{tabular}[c]{@{}c@{}}Use Case of\\ the Subject Systems\end{tabular}} \\ \hline
    
    \multirow{3}{*}{1} & \multirow{3}{*}{\begin{tabular}[c]{@{}c@{}}Compare each \\ MOSA with RS\end{tabular}} & \multirow{9}{*}{\begin{tabular}[c]{@{}c@{}}Problems \\ 1-10\\ (Table\ref{table:tabFitnessFunction})\end{tabular}} & \multirow{5}{*}{\begin{tabular}[c]{@{}c@{}}\\10, 20\\ 30, 40\\ 50, 60\\ 70, 80\\ 90, 100\end{tabular}} & \multicolumn{2}{l|}{\multirow{4}{*}{\begin{tabular}[c]{@{}l@{}}RS (only for RQ1)\\ EA: NSGA-II\\\phantom{EA:} MOCell\\\phantom{EA:} SPEA2\\ HA: CellDE\end{tabular}}} & \multirow{5}{*}{\begin{tabular}[c]{@{}c@{}}HV\\\rule{0pt}{3ex}\revision{IGD}\end{tabular}} & \multirow{9}{*}{\footnotesize\begin{tabular}[c]{@{}l@{}}Comparison Analysis:\\ - Kruskal-Wallis Test\\ - Mann-Whitney U Test\\ - Vargha and Delaney \Atwelve\\
    \\ Correlation Analysis:\\ - Spearman's Rank Correlation \\ \ \ Coefficient \\ \\ \revision{Multiple Hypothesis Testing:} \\ \revision{- Holm–Bonferroni Method}
    \end{tabular}} & \multirow{9}{*}{\begin{tabular}[c]{@{}c@{}}AW (AW1, AW1,\\ AW3, AW4)\\ \\ GS (GS1)\end{tabular}} \\
    & & & & \multicolumn{2}{l|}{} & & & \\
    & & & & \multicolumn{2}{l|}{} & & & \\ \cline{1-2}
    2 & \begin{tabular}[c]{@{}c@{}}Compare among\\ MOSA\end{tabular} & & & \multicolumn{2}{l|}{} & & & \\ \cline{1-2} \cline{5-7}
    3 & \begin{tabular}[c]{@{}c@{}}Study impacts of\\ TB for each SUT\end{tabular} & & & \multicolumn{2}{c|}{Best \MOSA} & ANUO & & \\ \cline{1-2} \cline{4-7}
    4 & \begin{tabular}[c]{@{}c@{}}Compare problems\\ for each SUT\end{tabular} & & Best TB & \multicolumn{2}{c|}{Best \MOSA} & ANUO & & \\ \hline
\end{tabular}

%% file: tables/tabStatistics.tex
\begin{tabular}{|c|c|c|c|}
    \hline
    \textbf{Subject System} & \textbf{Use Case} & \textbf{Execution Time (s)} & \textbf{\# of Test Cases}\\
    \hline
    
    \multirow{4}{*}{AW} & AW1 & 7924 & 420\\ \cline{2-4}
    & AW2 & 15250 & 776\\ \cline{2-4}
    & AW3 & 567960 & 857\\ \cline{2-4}
    & AW4 & 1655 & 296\\ \hline
    GS & GS1 & 118755 & 1799\\ \hline
    
\end{tabular}

%% file: tables/tabParameterSetting_new.tex
\begin{tabular}{|l|p{2.2in}|}
    \hline
    \textbf{Parameter} & \textbf{Settings} \\
    \hline
    
    Population Size & All: 100 \\ \hline
    Neighborhood & \begin{tabular}[l]{@{}l@{}}MOCell and CellDE: 1-hop neighbors \\(8 surrounding solutions)\end{tabular} \\ \hline
    Parents Selection & All: Binary Tournament \\ \hline
    Crossover  & \begin{tabular}[l]{@{}l@{}} \revision{All but CellDE: Simulated Binary},\\ \revision{CellDE: Differential evolution} \\crossover rate: 0.9\end{tabular} \\ \hline
    Mutation & \begin{tabular}[l]{@{}l@{}}All but CellDE: Polynomial, \\mutation rate: 1/n\end{tabular} \\ \hline
    Archive Size & MOCell and CellDE: 100 \\ \hline
    Max Generation & All: 25000 \\ \hline
    Times of Run & All: 100 \\ \hline
    
\end{tabular}

%% file: tables/algRank.tex
%
\begin{minipage}[t]{0.45\textwidth}
\begin{algorithm}[H]
\renewcommand{\algorithmicrequire}{\textbf{Input:}}
\renewcommand{\algorithmicensure}{\textbf{Output:}}
\caption{Rank Algorithm}\label{alg:RankAlgorithm}
\begin{algorithmic}[1]
\Require $algos[\,], len(algos) \geq 2$
\Ensure $algos[\,], rank[\,]$  $//$rank[i] is the rank value of algos[i]
\State $n \gets len(algos)$
\For{$i \gets 1$ to $n-1$}
    \For{$j \gets i+1$ to $n$}  $//$sort algos[]
        \If{$better(algos[i], algos[j])$}
            \State $switch(algos, i, j)$
        \EndIf
    \EndFor
\EndFor
\State $rank[1] \gets 1$
\For{$i \gets 2$ to $n$}  $//$set rank values for algos[]
    \If{$better(algos[i], algos[i-1])$}
        \State $rank[i] \gets rank[i-1]+1$
    \Else
        \State $rank[i] \gets rank[i-1]$
    \EndIf
\EndFor
\end{algorithmic}
\end{algorithm}
\vspace{-0.75cm}
\begin{threeparttable}
\begin{tablenotes}
\scriptsize \item[*] Function $better(algo1, algo2)$ compares $algo1$ with $algo2$, which returns the best algorithm based on these two conditions: 1) for \emph{PET}, p-value $<$ 0.05 and \Atwelve $<$ 0.5; 2) p-value $<$ 0.05 and \Atwelve $>$ 0.5.
\end{tablenotes}
\end{threeparttable}
\end{minipage}

%% file: tables/tabBestMOSA_HV_IGD_comb.tex
\begin{threeparttable}
\begin{tabular}{p{0.01in}p{0.47in}|p{0.28in}|p{0.35in}|p{0.28in}|p{0.28in}|p{0.28in}|p{0.28in}|p{0.28in}|p{0.28in}|p{0.28in}|p{0.34in}|}
    \cline{3-12}
    & & \multicolumn{10}{c|}{\textbf{Time Budgets}} \\ \cline{3-12}
    & & TB10 & TB20 & TB30 & TB40 & TB50 & TB60 & TB70 &TB80 & TB90 & TB100 \\ \hline 
    \multicolumn{1}{|p{0.07in}|}{\multirow{11}{*}{\textbf{\rotatebox{90}{Problems}}}} & Prob. 1 & 
    \cellcolor{blueS}\textbf{S} & \cellcolor{blueS}\textbf{S} & \cellcolor{blueS}\textbf{S} & \cellcolor{blueS}\textbf{S} & \cellcolor{blueS}\textbf{S} & \cellcolor{blueS}\textbf{S} & \cellcolor{blueS}\textbf{S} & \cellcolor{blueS}\textbf{S} & \cellcolor{blueS}\textbf{S} & \cellcolor{blueS}\textbf{S} \\ \cline{2-12}
    \multicolumn{1}{|p{0.07in}|}{} & Prob. 2 & \cellcolor{blueS}\textbf{S} & \cellcolor{blueS}\textbf{S} & \cellcolor{blueS}\textbf{S} & \cellcolor{blueS}\textbf{S} & \cellcolor{blueS}\textbf{S} & \cellcolor{blueS}\textbf{S} & \cellcolor{blueS}\textbf{S} & \cellcolor{blueS}\textbf{S} & \cellcolor{blueS}\textbf{S} & \cellcolor{blueS}\textbf{S} \\ \cline{2-12}
    \multicolumn{1}{|p{0.07in}|}{} & Prob. 3 & \cellcolor{multi}\textbf{M/S} & \cellcolor{blueS}\textbf{S} & \cellcolor{blueS}\textbf{S} & \cellcolor{blueS}\textbf{S} & \cellcolor{blueS}\textbf{S} & \cellcolor{blueS}\textbf{S} & \cellcolor{blueS}\textbf{S} & \cellcolor{blueS}\textbf{S} & \cellcolor{blueS}\textbf{S} & \cellcolor{blueS}\textbf{S} \\ \cline{2-12}
    \multicolumn{1}{|p{0.07in}|}{} & Prob. 4 & \cellcolor{blueS}\textbf{S} & \cellcolor{blueS}\textbf{S} & \cellcolor{blueS}\textbf{S} & \cellcolor{blueS}\textbf{S} & \cellcolor{blueS}\textbf{S} & \cellcolor{blueS}\textbf{S} & \cellcolor{blueS}\textbf{S} & \cellcolor{blueS}\textbf{S} & \cellcolor{blueS}\textbf{S} & \cellcolor{blueS}\textbf{S} \\ \cline{2-12}
    \multicolumn{1}{|p{0.07in}|}{} & Prob. 5 & \cellcolor{blueS}\textbf{S} & \cellcolor{blueS}\textbf{S} & \cellcolor{blueS}\textbf{S} & \cellcolor{blueS}\textbf{S} & \cellcolor{blueS}\textbf{S} & \cellcolor{blueS}\textbf{S} & \cellcolor{blueS}\textbf{S} & \cellcolor{blueS}\textbf{S} & \cellcolor{blueS}\textbf{S} & \cellcolor{blueS}\textbf{S} \\ \cline{2-12}
    \multicolumn{1}{|p{0.07in}|}{} & Prob. 6 & \cellcolor{greenM}\textbf{M} & \cellcolor{greenM}\textbf{M} & \cellcolor{greenM}\textbf{M} & \cellcolor{blueS}\textbf{S} & \cellcolor{multi}\textbf{M/S} & \cellcolor{multi}\textbf{M/S} & \cellcolor{blueS}\textbf{S} & \cellcolor{blueS}\textbf{S} & \cellcolor{blueS}\textbf{S} & \cellcolor{blueS}\textbf{S} \\ \cline{2-12}
    \multicolumn{1}{|p{0.07in}|}{} & Prob. 7 & \cellcolor{blueS}\textbf{S} & \cellcolor{blueS}\textbf{S} & \cellcolor{blueS}\textbf{S} & \cellcolor{blueS}\textbf{S} & \cellcolor{blueS}\textbf{S} & \cellcolor{blueS}\textbf{S} & \cellcolor{blueS}\textbf{S} & \cellcolor{blueS}\textbf{S} & \cellcolor{blueS}\textbf{S} & \cellcolor{blueS}\textbf{S} \\ \cline{2-12}
    \multicolumn{1}{|p{0.07in}|}{} & Prob. 8 & \cellcolor{multi}\textbf{M/S} & \cellcolor{multi}\textbf{M/S} & \cellcolor{blueS}\textbf{S} & \cellcolor{blueS}\textbf{S} & \cellcolor{blueS}\textbf{S} & \cellcolor{blueS}\textbf{S} & \cellcolor{blueS}\textbf{S} & \cellcolor{blueS}\textbf{S} & \cellcolor{blueS}\textbf{S} & \cellcolor{blueS}\textbf{S} \\ \cline{2-12}
    \multicolumn{1}{|p{0.07in}|}{} & Prob. 9 & \cellcolor{blueS}\textbf{S} & \cellcolor{blueS}\textbf{S} & \cellcolor{blueS}\textbf{S} & \cellcolor{blueS}\textbf{S} & \cellcolor{blueS}\textbf{S} & \cellcolor{blueS}\textbf{S} & \cellcolor{blueS}\textbf{S} & \cellcolor{blueS}\textbf{S} & \cellcolor{blueS}\textbf{S} & \cellcolor{blueS}\textbf{S} \\ \cline{2-12}
    \multicolumn{1}{|p{0.07in}|}{} & Prob. 10 & \cellcolor{blueS}\textbf{S} & \cellcolor{multi}\textbf{N/M} & \cellcolor{blueS}\textbf{S} & \cellcolor{blueS}\textbf{S} & \cellcolor{blueS}\textbf{S} & \cellcolor{blueS}\textbf{S} & \cellcolor{blueS}\textbf{S} & \cellcolor{blueS}\textbf{S} & \cellcolor{blueS}\textbf{S} & \cellcolor{blueS}\textbf{S} \\ \hline 
\end{tabular}

\begin{tablenotes}
\item[*] Note that \textbf{N:} NSGA-II, \textbf{M:} MOCell, and \textbf{S:} SPEA2
\end{tablenotes}

\end{threeparttable}

%% file: tables/tabSpearm_casestudies_HV_IGD.tex
\begin{tabular}{lrrrrrr}
    
    \toprule
    \textbf{Problem} & \textbf{AW1}  & \textbf{AW2}  & \textbf{AW3} & \textbf{AW4}  & \textbf{GS1}  \\
    \midrule
    \probFirst        & 0.58  & 0.93  & -0.25  & 0.99  & 0.96   \\
    \probSecond        & 0.96    & 0.94  & -0.12   & 0.99  & 0.99  \\
    \probThird        & 0.96   & 0.98  & -0.95  & 0.04 & -0.06  \\
    \probFourth       & 0.96   & 0.94  & -0.18   & 0.99   & 0.98  \\
    \probFifth   & 0.41   & 0.89  & -0.21  & 0.99  & 0.96  \\
    \probSixth    & 0.26  & 0.45  & -0.52  & -0.01& -0.06\\
    \probSeventh  & 0.50  & 0.89   & -0.20  & 0.99  & 0.95  \\
    \probEighth    & 0.96   & 0.97  & -0.96  & 0.07 & \underline{0.00} \\
    \probNineth   & 0.96   & 0.94  & -0.09   & 0.99  & 0.98  \\
    \probTenth   & 0.96   & 0.98  & -0.95  & 0.03 & -0.08\\
    \bottomrule

\end{tabular}

%% file: tables/tabBestTBtoANOU_HV_IGD.tex
\begin{tabular}{lrrrrr}
    
    \toprule
    \textbf{Problem} & \textbf{AW1} & \textbf{AW2} & \textbf{AW3} & \textbf{AW4} & \textbf{GS1}  \\
    \midrule
    \probFirst & \textbf{TB100} & \textbf{TB100} & TB20 & \textbf{TB100} & \textbf{TB100} \\
    \probSecond & \textbf{TB100} & \textbf{TB100} & TB10& \textbf{TB100} & \textbf{TB100} \\
    \probThird & \textbf{TB100} & \textbf{TB100} & TB10 & TB80 & TB20 \\
    \probFourth & \textbf{TB100} & \textbf{TB100} & TB10 & \textbf{TB100} & \textbf{TB100} \\
    \probFifth & \textbf{TB100} & \textbf{TB100} & TB20 & \textbf{TB100} & \textbf{TB100} \\
    \probSixth & \textbf{TB100} & \textbf{TB100} & {TB30} & TB60 & TB50 \\
    \probSeventh & \textbf{TB100} & \textbf{TB100} & TB20 & \textbf{TB100} & \textbf{TB100} \\
    \probEighth & \textbf{TB100} & \textbf{TB100} & TB10 & TB90 & TB80 \\
    \probNineth & \textbf{TB100} & \textbf{TB100} & TB10 & \textbf{TB100} & \textbf{TB100} \\
    \probTenth & \textbf{TB100} & \textbf{TB100} & TB10 & {TB60} & TB40 \\
    \bottomrule

\end{tabular}

%% file: tables/et.tex
\begin{tabular}{ l r r r r r } 
\toprule 
\textbf{Use Case} & \textbf{Avg} & \textbf{Max} & \textbf{Min} & $\bm{\sigma}$ & \textbf{Avg(ET/NOU)}\\ 
\midrule 
\emph{AW1} & 18.87 & 300.01 & 4.82 & 42.60 & 9.44  \\ 
\emph{AW2} & 19.65 & 300.00 & 10.25 & 37.01 & 11.20  \\ 
\emph{AW3} & 662.73 & 1200.02 & 373.05 & 165.87 & 117.40  \\ 
\emph{AW4} & 5.59 & 6.66 & 1.78 & 1.26 & 0.58  \\ 
\emph{GS1} & 66.01 & 99.70 & 10.30 & 14.66 & 23.50  \\ 
\bottomrule
\end{tabular} 

%% file: tables/spearman_nu_nou.tex
\begin{tabular}{ c  r r r }
\toprule 
\textbf{Use Case}  & $\bm{\rho}$ & $\bm{p}$\textbf{-value} & \textbf{Avg(ET/NU)} \\ 
\midrule 
\emph{AW1} & \underline{0.02} & \underline{0.67} & 2.51  \\ 
\emph{AW2} & \underline{0.03} & \underline{0.45} & 3.18  \\
\emph{AW3} & 0.44 & $<$0.01 & 69.75  \\ 
\emph{AW4} & 0.86 & $<$0.01 & 0.43  \\ 
\emph{GS1} & 0.39 & $<$0.01 & 8.54  \\ 
\bottomrule 
\end{tabular} 

%% file: tables/tabProblemRank_comb.tex
\begin{threeparttable}
\begin{tabular}{|p{0.3in}|p{0.2in}|p{0.2in}|p{0.2in}|p{0.2in}|p{0.2in}|p{0.2in}|p{0.2in}|p{0.2in}|p{0.2in}|p{0.2in}|}
    \hline 
    \multirow{2}{*}{\textbf{UC}} & \multicolumn{10}{p{2.0in}|}{\textbf{Problem}}\\ \cline{2-11}
    & \textbf{1} & \textbf{2} & \textbf{3} & \textbf{4} & \textbf{5} & \textbf{6} & \textbf{7} & \textbf{8} & \textbf{9} & \textbf{10} \\ \hline
    
    AW1 & 8 & 1 & 4 & 1 & 7 & \cellcolor[gray]{.8}{10} & 9 & 5 & 1 & 6 \\ \cline{1-11}
    AW2 & 4 & 1 & 7 & 1 & 6 & \cellcolor[gray]{.8}{10} & 4 & {8} & 1 & 8 \\ \cline{1-11}
    AW3 & 5 & 2 & \cellcolor[gray]{.8}{9} & 1 & 4 & {7} & 5 & {8} & 2 & \cellcolor[gray]{.8}{9} \\ \cline{1-11}
    AW4 & 4 & 2 &{8} & 3 & 4 & \cellcolor[gray]{.8}{10} & 4 & {7} & 1 & {9} \\ \cline{1-11}
    GS1 & 5 & 2 & {7} & 1 & 4 & \cellcolor[gray]{.8}{10} & 5 & {8} & 3 & {9} \\ \cline{1-11}
    \textit{}{All} & 26 & 8 & {35} & 7 & 25 & \cellcolor[gray]{.8}{47} & 27 & {36} & 8 & {41} \\ \hline
\end{tabular}
\end{threeparttable}

%% file: tables/tabProblemConfidence_comb.tex
\begin{threeparttable}
\begin{tabular}{|p{0.3in}|p{0.2in}|p{0.2in}|p{0.2in}|p{0.2in}|p{0.2in}|p{0.2in}|p{0.2in}|p{0.2in}|p{0.2in}|p{0.2in}|}
    \hline 
    \multirow{2}{*}{\textbf{UC}} & \multicolumn{10}{p{2.0in}|}{\textbf{Problem}}\\ \cline{2-11}
    & \textbf{1} & \textbf{2} & \textbf{3} & \textbf{4} & \textbf{5} & \textbf{6} & \textbf{7} & \textbf{8} & \textbf{9} & \textbf{10} \\ \hline
    
    AW1 & {15\%}  & 2\% & 8\% & 2\% & {13\%} & \cellcolor[gray]{.8}{19\%} & {17\%} & 10\% & 2\% & 12\% \\ \cline{1-11}
    AW2 & 8\% & 2\% &{14\%} & 2\%  & 12\% & \cellcolor[gray]{.8}{20\%} & 8\%  & {16\%} & 2\%  & {16\%} \\ \cline{1-11}
    AW3 & 10\% & 4\% & \cellcolor[gray]{.8}{17\% }& 2\% & 8\% & {13\%} & 10\%  & {15\%}  & 4\%  & \cellcolor[gray]{.8}{17\% }\\ \cline{1-11}
    AW4 & 8\%  & 4\% & {15\%} & 6\% & 8\%  & \cellcolor[gray]{.8}{19\%} & 8\% & {13\%}  & 2\%  & {17\%} \\ \cline{1-11}
    GS1 & 9\% & 4\% & {13\%} & 2\% & 7\%  & \cellcolor[gray]{.8}{19\%} & 9\%  & {15\%} & 6\%  & {17\%} \\ \cline{1-11}
    \textit{All} & 10\% & 3\% &{13\%}& 3\% & 10\% & \cellcolor[gray]{.8}{18\%} & 10\% & {14\%} & 3\%  & {16\%} \\ \hline
\end{tabular}
\end{threeparttable}

%% file: tables/spearman_um_nu_nou.tex
\begin{tabular}{ c r r r r r r r r}\\ 
\toprule 
\multirow{2}{*}{\textbf{Use Case}}   & \multicolumn{2}{c}{\textbf{\nou$\bm{\sim}$\um}}& \multicolumn{2}{c}{\textbf{\nou$\bm{\sim}$\nsu}} & \multicolumn{2}{c}{\textbf{\nou$\bm{\sim}$\nsuu}} & \multicolumn{2}{c}{\textbf{\nou$\bm{\sim}$\usp}} \\ 
&$\bm{\rho}$ & $\bm{p}$\textbf{-value} & $\bm{\rho}$ & $\bm{p}$\textbf{-value} & $\bm{\rho}$ & $\bm{p}$\textbf{-value}& $\bm{\rho}$ & $\bm{p}$\textbf{-value}\\ 
\midrule 
\emph{AW1} &  0.02  &  0.74 &  0.02  &  0.67 &  -0.07  &  0.15 & -0.09 & 0.06 \\ 
\emph{AW2} &  -0.01  &  0.70 &  0.03  &  0.45 &  -0.07  &  0.04 & -0.03 & 0.46 \\ 
\emph{AW3} &  0.06  &  0.10 &  0.44  &  $<$0.01 &  0.25  &  $<$0.01 & 0.17 & $<$0.01 \\ 
\emph{AW4} &  -0.05  &  0.41 &  0.86  &  $<$0.01 &  0.57  &  $<$0.01 & 0.58 & $<$0.01 \\ 
\emph{GS1} &  0.11  &  $<$0.01 &  0.39  &  $<$0.01 &  0.28  &  $<$0.01 & 0.19 & $<$0.01 \\ 
\midrule 
  & \multicolumn{2}{c}{\textbf{\um$\bm{\sim}$\nsu}}& \multicolumn{2}{c}{\textbf{\um$\bm{\sim}$\nsuu}} & \multicolumn{2}{c}{\textbf{\um$\bm{\sim}$\usp}} & \multicolumn{2}{c}{} \\ 
\midrule 
\emph{AW1} &  -0.12  &  0.01 &  -0.16  &  $<$0.01 &  -0.13  &  $<$0.01 &  &  \\ 
\emph{AW2} &  -0.16  &  $<$0.01 &  -0.29  &  $<$0.01 &  -0.16  &  $<$0.01 &  &  \\ 
\emph{AW3} &  -0.28  &  $<$0.01 &  -0.43  &  $<$0.01 &  -0.20  &  $<$0.01 &  &  \\ 
\emph{AW4} &  -0.13  &  0.02 &  -0.35  &  $<$0.01 &  -0.29  &  $<$0.01 &  &  \\ 
\emph{GS1} &  -0.25  &  $<$0.01 &  -0.53  &  $<$0.01 &  -0.12  &  $<$0.01 &  &  \\ 
\midrule 
 & \multicolumn{2}{c}{\textbf{\nsu$\bm{\sim}$\nsuu}} & \multicolumn{2}{c}{\textbf{\nsu$\bm{\sim}$\usp}}& \multicolumn{2}{c}{\textbf{\nsuu$\bm{\sim}$\usp}}  & \multicolumn{2}{c}{} \\ 
\midrule 
\emph{AW1} &  0.36  &  $<$0.01 &  0.30  &  $<$0.01 &  0.80  &  $<$0.01 &  &  \\ 
\emph{AW2} &  0.37  &  $<$0.01 &  0.22  &  $<$0.01 &  0.77  &  $<$0.01 &  &  \\ 
\emph{AW3} &  0.40  &  $<$0.01 &  0.24  &  $<$0.01 &  0.77  &  $<$0.01 &  &  \\ 
\emph{AW4} &  0.57  &  $<$0.01 &  0.58  &  $<$0.01 &  1.00  &  $<$0.01 &  &  \\ 
\emph{GS1} &  0.47  &  $<$0.01 &  0.27  &  $<$0.01 &  0.28  &  $<$0.01 &  &  \\ 
\bottomrule 
\end{tabular} 

%% file: main.bbl
\begin{thebibliography}{10}
\providecommand{\url}[1]{#1}
\csname url@samestyle\endcsname
\providecommand{\newblock}{\relax}
\providecommand{\bibinfo}[2]{#2}
\providecommand{\BIBentrySTDinterwordspacing}{\spaceskip=0pt\relax}
\providecommand{\BIBentryALTinterwordstretchfactor}{4}
\providecommand{\BIBentryALTinterwordspacing}{\spaceskip=\fontdimen2\font plus
\BIBentryALTinterwordstretchfactor\fontdimen3\font minus \fontdimen4\font\relax}
\providecommand{\BIBforeignlanguage}[2]{{%
\expandafter\ifx\csname l@#1\endcsname\relax
\typeout{** WARNING: IEEEtran.bst: No hyphenation pattern has been}%
\typeout{** loaded for the language `#1'. Using the pattern for}%
\typeout{** the default language instead.}%
\else
\language=\csname l@#1\endcsname
\fi
#2}}
\providecommand{\BIBdecl}{\relax}
\BIBdecl

\bibitem{patton2006software}
R.~Patton, \emph{Software testing}.\hskip 1em plus 0.5em minus 0.4em\relax Pearson Education India, 2006.

\bibitem{wang2016enhancing}
S.~Wang, S.~Ali, T.~Yue, {\O}.~Bakkeli, and M.~Liaaen, ``Enhancing test case prioritization in an industrial setting with resource awareness and multi-objective search,'' in \emph{Proceedings of the 38th International Conference on Software Engineering Companion}, 2016, pp. 182--191.

\bibitem{Resource-allocationSBST}
R.~Pietrantuono and S.~Russo, ``Search-based optimization for the testing resource allocation problem: Research trends and opportunities,'' in \emph{2018 IEEE/ACM 11th International Workshop on Search-Based Software Testing (SBST)}, 2018, pp. 6--12.

\bibitem{wang2018oracles}
C.~Wang, F.~Pastore, and L.~Briand, ``Oracles for testing software timeliness with uncertainty,'' \emph{ACM Transactions on Software Engineering and Methodology (TOSEM)}, vol.~28, no.~1, pp. 1--30, 2018.

\bibitem{shin2018test}
S.~Y. Shin, S.~Nejati, M.~Sabetzadeh, L.~C. Briand, and F.~Zimmer, ``Test case prioritization for acceptance testing of cyber physical systems: a multi-objective search-based approach,'' in \emph{Proceedings of the 27th ACM SIGSOFT International Symposium on Software Testing and Analysis}, 2018, pp. 49--60.

\bibitem{menghi2019generating}
C.~Menghi, S.~Nejati, K.~Gaaloul, and L.~C. Briand, ``Generating automated and online test oracles for simulink models with continuous and uncertain behaviors,'' in \emph{Proceedings of the 2019 27th acm joint meeting on european software engineering conference and symposium on the foundations of software engineering}, 2019, pp. 27--38.

\bibitem{shin2021uncertainty}
S.~Y. Shin, K.~Chaouch, S.~Nejati, M.~Sabetzadeh, L.~C. Briand, and F.~Zimmer, ``Uncertainty-aware specification and analysis for hardware-in-the-loop testing of cyber-physical systems,'' \emph{Journal of Systems and Software}, vol. 171, p. 110813, 2021.

\bibitem{camilli2021uncertainty}
M.~Camilli, A.~Gargantini, P.~Scandurra, and C.~Trubiani, ``Uncertainty-aware exploration in model-based testing,'' in \emph{2021 14th IEEE Conference on Software Testing, Verification and Validation (ICST)}.\hskip 1em plus 0.5em minus 0.4em\relax IEEE, 2021, pp. 71--81.

\bibitem{zhang2019uncertainty}
M.~Zhang, S.~Ali, and T.~Yue, ``Uncertainty-wise test case generation and minimization for cyber-physical systems,'' \emph{Journal of Systems and Software}, vol. 153, pp. 1--21, 2019.

\bibitem{liu2007uncertainty}
B.~Liu, ``Uncertainty theory,'' in \emph{Uncertainty theory}.\hskip 1em plus 0.5em minus 0.4em\relax Springer, 2007, pp. 205--234.

\bibitem{deb2002fast}
K.~Deb, A.~Pratap, S.~Agarwal, and T.~Meyarivan, ``A fast and elitist multiobjective genetic algorithm: Nsga-ii,'' \emph{IEEE transactions on evolutionary computation}, vol.~6, no.~2, pp. 182--197, 2002.

\bibitem{nebro2007design}
A.~J. Nebro, J.~J. Durillo, F.~Luna, B.~Dorronsoro, and E.~Alba, ``Design issues in a multiobjective cellular genetic algorithm,'' in \emph{International Conference on Evolutionary Multi-Criterion Optimization}.\hskip 1em plus 0.5em minus 0.4em\relax Springer, 2007, pp. 126--140.

\bibitem{nebro2009mocell}
------, ``Mocell: A cellular genetic algorithm for multiobjective optimization,'' \emph{International Journal of Intelligent Systems}, vol.~24, no.~7, pp. 726--746, 2009.

\bibitem{zitzler2001spea2}
E.~Zitzler, M.~Laumanns, and L.~Thiele, ``Spea2: Improving the strength pareto evolutionary algorithm,'' \emph{TIK-report}, vol. 103, 2001.

\bibitem{durillo2008solving}
J.~J. Durillo, A.~J. Nebro, F.~Luna, and E.~Alba, ``Solving three-objective optimization problems using a new hybrid cellular genetic algorithm,'' in \emph{International Conference on Parallel Problem Solving from Nature}.\hskip 1em plus 0.5em minus 0.4em\relax Springer, 2008, pp. 661--670.

\bibitem{zhang2016understanding}
M.~Zhang, B.~Selic, S.~Ali, T.~Yue, O.~Okariz, and R.~Norgren, ``Understanding uncertainty in cyber-physical systems: a conceptual model,'' in \emph{Modelling Foundations and Applications: 12th European Conference, ECMFA 2016, Held as Part of STAF 2016, Vienna, Austria, July 6-7, 2016, Proceedings 12}.\hskip 1em plus 0.5em minus 0.4em\relax Springer, 2016, pp. 247--264.

\bibitem{zhang2019uncertaintywise}
M.~Zhang, S.~Ali, T.~Yue, R.~Norgren, and O.~Okariz, ``Uncertainty-wise cyber-physical system test modeling,'' \emph{Software \& Systems Modeling}, vol.~18, no.~2, pp. 1379--1418, 2019.

\bibitem{zhang2022test}
Q.~Zhang, C.~Fang, W.~Sun, S.~Yu, Y.~Xu, and Y.~Liu, ``Test case prioritization using partial attention,'' \emph{Journal of Systems and Software}, vol. 192, p. 111419, 2022.

\bibitem{liu2012there}
B.~Liu, ``Why is there a need for uncertainty theory,'' \emph{Journal of Uncertain Systems}, vol.~6, no.~1, pp. 3--10, 2012.

\bibitem{walter1987real}
R.~Walter, ``Real and complex analysis,'' 1987.

\bibitem{jMetal}
A.~J. Nebro and J.~J. Durillo, ``jmetal,'' \url{http://jmetal.sourceforge.net/}, 2016.

\bibitem{li2017zen}
Y.~Li, T.~Yue, S.~Ali, and L.~Zhang, ``Zen-reqoptimizer: a search-based approach for requirements assignment optimization,'' \emph{Empirical Software Engineering}, vol.~22, no.~1, pp. 175--234, 2017.

\bibitem{yue2014applying}
T.~Yue and S.~Ali, ``Applying search algorithms for optimizing stakeholders familiarity and balancing workload in requirements assignment,'' in \emph{Proceedings of the 2014 Annual Conference on Genetic and Evolutionary Computation}, 2014, pp. 1295--1302.

\bibitem{lu2016nonconformity}
H.~Lu, T.~Yue, S.~Ali, and L.~Zhang, ``Nonconformity resolving recommendations for product line configuration,'' in \emph{2016 IEEE International Conference on Software Testing, Verification and Validation (ICST)}.\hskip 1em plus 0.5em minus 0.4em\relax IEEE, 2016, pp. 57--68.

\bibitem{wang2014multi}
S.~Wang, D.~Buchmann, S.~Ali, A.~Gotlieb, D.~Pradhan, and M.~Liaaen, ``Multi-objective test prioritization in software product line testing: an industrial case study,'' in \emph{Proceedings of the 18th International Software Product Line Conference-Volume 1}, 2014, pp. 32--41.

\bibitem{U-Test}
U-Test, ``Use cases - industrial case studies,'' \url{http://www.u-test.eu/use-cases/}, 2017.

\bibitem{ali2020quality}
S.~Ali, P.~Arcaini, D.~Pradhan, S.~A. Safdar, and T.~Yue, ``Quality indicators in search-based software engineering: An empirical evaluation,'' \emph{ACM Transactions on Software Engineering and Methodology (TOSEM)}, vol.~29, no.~2, pp. 1--29, 2020.

\bibitem{dodge2008concise}
Y.~Dodge, \emph{The concise encyclopedia of statistics}.\hskip 1em plus 0.5em minus 0.4em\relax Springer Science \& Business Media, 2008.

\bibitem{arcuri2011practical}
A.~Arcuri and L.~Briand, ``A practical guide for using statistical tests to assess randomized algorithms in software engineering,'' in \emph{2011 33rd International Conference on Software Engineering (ICSE)}.\hskip 1em plus 0.5em minus 0.4em\relax IEEE, 2011, pp. 1--10.

\bibitem{holm1979simple}
S.~Holm, ``A simple sequentially rejective multiple test procedure,'' \emph{Scandinavian journal of statistics}, pp. 65--70, 1979.

\bibitem{spearman1987proof}
C.~Spearman, ``The proof and measurement of association between two things,'' \emph{The American journal of psychology}, vol. 100, no. 3/4, pp. 441--471, 1987.

\bibitem{schober2018correlation}
P.~Schober, C.~Boer, and L.~A. Schwarte, ``Correlation coefficients: appropriate use and interpretation,'' \emph{Anesthesia \& Analgesia}, vol. 126, no.~5, pp. 1763--1768, 2018.

\bibitem{sheskin2003handbook}
D.~J. Sheskin, \emph{Handbook of parametric and nonparametric statistical procedures}.\hskip 1em plus 0.5em minus 0.4em\relax Chapman and Hall/CRC, 2003.

\bibitem{wang2013minimizing}
S.~Wang, S.~Ali, and A.~Gotlieb, ``Minimizing test suites in software product lines using weight-based genetic algorithms,'' in \emph{Proceedings of the 15th annual conference on Genetic and evolutionary computation}, 2013, pp. 1493--1500.

\bibitem{yue2013facilitating}
T.~Yue, L.~C. Briand, and Y.~Labiche, ``Facilitating the transition from use case models to analysis models: Approach and experiments,'' \emph{ACM Transactions on Software Engineering and Methodology (TOSEM)}, vol.~22, no.~1, pp. 1--38, 2013.

\bibitem{lu2019ant}
C.~Lu, J.~Zhong, Y.~Xue, L.~Feng, and J.~Zhang, ``Ant colony system with sorting-based local search for coverage-based test case prioritization,'' \emph{IEEE Transactions on Reliability}, vol.~69, no.~3, pp. 1004--1020, 2019.

\bibitem{marijan2017titan}
D.~Marijan, M.~Liaaen, A.~Gotlieb, S.~Sen, and C.~Ieva, ``Titan: Test suite optimization for highly configurable software,'' in \emph{2017 IEEE International Conference on Software Testing, Verification and Validation (ICST)}.\hskip 1em plus 0.5em minus 0.4em\relax IEEE, 2017, pp. 524--531.

\bibitem{hierons2020many}
R.~M. Hierons, M.~Li, X.~Liu, J.~A. Parejo, S.~Segura, and X.~Yao, ``Many-objective test suite generation for software product lines,'' \emph{ACM Transactions on Software Engineering and Methodology (TOSEM)}, vol.~29, no.~1, pp. 1--46, 2020.

\bibitem{parejo2016multi}
J.~A. Parejo, A.~B. S{\'a}nchez, S.~Segura, A.~Ruiz-Cort{\'e}s, R.~E. Lopez-Herrejon, and A.~Egyed, ``Multi-objective test case prioritization in highly configurable systems: A case study,'' \emph{Journal of Systems and Software}, vol. 122, pp. 287--310, 2016.

\bibitem{wang2015cost}
S.~Wang, S.~Ali, and A.~Gotlieb, ``Cost-effective test suite minimization in product lines using search techniques,'' \emph{Journal of Systems and Software}, vol. 103, pp. 370--391, 2015.

\bibitem{wang2016practical}
S.~Wang, S.~Ali, T.~Yue, Y.~Li, and M.~Liaaen, ``A practical guide to select quality indicators for assessing pareto-based search algorithms in search-based software engineering,'' in \emph{Proceedings of the 38th International Conference on Software Engineering}, 2016, pp. 631--642.

\bibitem{arrieta2019search}
A.~Arrieta, S.~Wang, G.~Sagardui, and L.~Etxeberria, ``Search-based test case prioritization for simulation-based testing of cyber-physical system product lines,'' \emph{Journal of Systems and Software}, vol. 149, pp. 1--34, 2019.

\bibitem{arrieta2016test}
------, ``Test case prioritization of configurable cyber-physical systems with weight-based search algorithms,'' in \emph{Proceedings of the Genetic and Evolutionary Computation Conference 2016}, 2016, pp. 1053--1060.

\bibitem{arrieta2017employing}
A.~Arrieta, S.~Wang, U.~Markiegi, G.~Sagardui, and L.~Etxeberria, ``Employing multi-objective search to enhance reactive test case generation and prioritization for testing industrial cyber-physical systems,'' \emph{IEEE Transactions on Industrial Informatics}, vol.~14, no.~3, pp. 1055--1066, 2017.

\bibitem{zhang2020uncertainty}
H.~Zhang, M.~Zhang, T.~Yue, S.~Ali, and Y.~Li, ``Uncertainty-wise requirements prioritization with search,'' \emph{ACM Transactions on Software Engineering and Methodology (TOSEM)}, vol.~30, no.~1, pp. 1--54, 2020.

\end{thebibliography}
